\documentclass[english,11pt]{article}
\usepackage[T1]{fontenc}
\usepackage[latin1]{inputenc}
\usepackage[pdftex]{graphicx}
\usepackage{verbatim}
\usepackage[width=\textwidth]{caption}
\usepackage{amsmath}
\usepackage{a4wide}
\usepackage{amssymb}
\usepackage[english]{babel}
\usepackage[]{float}
\usepackage[]{placeins}
\usepackage{flafter}
\usepackage[]{longtable}
\usepackage{verbatim}
\usepackage{amsopn}
\usepackage{dsfont}
\usepackage{color}
\usepackage{array}
\usepackage{natbib}
\usepackage{xr}
\newtheorem{thm}{Theorem}[section]

\newtheorem{hyp}{Assumption}

\newcommand{\convL}{\stackrel{L}{\longrightarrow}}
\newcommand{\R}{\mathbb R}

\newcommand{\Supp}{\mathcal{S}}
\newcommand{\indep}{\perp \!\!\! \perp}
\newcommand{\eps}{\varepsilon}
\newcommand{\interior}{\overset{\circ}}

\newcommand{\Sig}{\sum_{i=1}^n}

\renewcommand{\citep}[1]{\citeauthor{#1}, \citeyear{#1}}
\newcolumntype{C}[1]{>{\centering\let\newline\\\arraybackslash\hspace{0pt}}m{#1}}


\date{\today}
\begin{document}

\title{Fuzzy Differences-in-Differences\thanks{This paper is a merged and revised version of
\cite{deChaisemartin14} and \cite{deChaisemartin13}. We thank Yannick Guyonvarch for outstanding research assistance, and are very grateful to Esther Duflo for sharing her data with us. We also want to thank the editor, five anonymous referees, Alberto Abadie, Joshua Angrist, Marc Gurgand, Guido Imbens, Rafael Lalive, Thierry Magnac, Blaise Melly, Roland Rathelot, Bernard Salanié, Frank Vella, Fabian Waldinger, Yichong Zhang,  and participants at various conferences and seminars for their helpful comments.}}

\author{Cl\'{e}ment de Chaisemartin%
\thanks{University of California at Santa Barbara, clementdechaisemartin@ucsb.edu%
} \and Xavier D'Haultf\oe{}uille%
\thanks{CREST, xavier.dhaultfoeuille@ensae.fr%
}} \maketitle ~\vspace{-1cm}
\begin{abstract}
Difference-in-differences (DID) is a method to evaluate the effect of a treatment. In its basic version, a ``control group'' is untreated at two dates, whereas a ``treatment group'' becomes fully treated at the second date. However, in many applications of the DID method, the treatment rate only increases more in the treatment group. In such fuzzy designs, a popular estimator of treatment effects is the DID of the outcome divided by the DID of the treatment. We show that this ratio identifies a local average treatment effect only if two homogeneous treatment effect assumptions are satisfied.
We then propose two alternative estimands that do not rely on any assumption on treatment effects, and that can be used when the treatment rate does not change over time in the control group. We prove that the corresponding estimators are asymptotically normal. Finally, we use our results to revisit Duflo (2001).
\end{abstract}



\newpage{}

\section{Introduction}

Difference-in-differences (DID) is a method to evaluate the effect of a treatment
in the absence of experimental data. In its basic version, a ``control group'' is untreated at two dates, whereas a
``treatment group'' becomes treated at the second date. If the trend on the outcome is the same in both groups, the
so-called common trends assumption, one can measure the effect of the treatment by
comparing the evolution of the outcome in the two groups.

\medskip
However, in many applications of the DID method the treatment rate increases more in
some groups than in others between the two dates, but there is no group that experiences a sharp change in treatment,
and there is also no group that remains fully untreated. In such fuzzy designs, a popular estimator
of treatment effects is the DID of the outcome divided by the DID of the treatment,
the so-called Wald-DID estimator. Other popular estimation methods are 2SLS or group-level OLS regressions with time and group
fixed effects. In \cite{deChaisemartin16}, we show that these regressions estimate a weighted average of Wald-DIDs across groups. We also show that 10.1\% of all papers published by the American Economic Review between 2010 and 2012 estimate either a simple Wald-DID, or the aforementioned IV or OLS regression. Despite its popularity, to our knowledge no paper has studied under which condition the Wald-DID estimand identifies a causal effect in a model with heterogeneous treatment effects.

\medskip
This paper makes the following contributions. Hereafter, let ``switchers'' refer to units that become treated at the second date. We start by showing that the Wald-DID estimand identifies the local average treatment effect (LATE) of treatment group switchers only if two treatment effect homogeneity assumptions are satisfied, on top of the usual common trend assumption. First, the LATE of units treated at both dates must not change over time. Second, 
when the share of treated units changes between the two dates in the control group, the LATEs of treatment and control group switchers must be equal. 
Then, we propose two alternative estimands of the same LATE. They do not rely on any treatment effect homogeneity assumption, and they can be used when the share of treated units is stable in the control group. The first one, the time-corrected Wald ratio (Wald-TC), relies on common trends assumptions within subgroups of units sharing the same treatment at the first date. The second one, the changes-in-changes Wald ratio (Wald-CIC), generalizes the changes-in-changes (CIC) estimand introduced by \cite{Athey06} to fuzzy designs. It relies on the assumption that a control and a treatment group unit with the same outcome and the same treatment at the first date will also have the same outcome at the second date.\footnote{Strictly speaking, the assumption underlying the CIC and Wald-CIC estimands is slightly weaker than what we describe. We still find this presentation helpful for pedagogical purposes.} Finally, we discuss how researchers can choose between the Wald-DID, Wald-TC, and Wald-CIC estimands.

\medskip
We then extend these identification results in several important directions. We start by showing how our results can be used in applications with more than two groups.
We also show that our results extend to applications with a non-binary treatment variable. Finally, we show that under the same assumptions as those underlying the Wald-TC and Wald-CIC estimands, the LATE of treatment group switchers is partially identified when the share of treated units changes over time in the control group. We also consider estimators of the Wald-DID, Wald-TC, and Wald-CIC, and we derive their limiting distributions.\footnote{A Stata package computing the estimators is available on the authors' webpages.}

\medskip
Finally, we use our results to revisit findings in \cite{Duflo01} on returns to education. Years of schooling increased substantially in the control group used by the author, so we show that her estimate heavily relies on the assumption that returns to schooling are the same in her two groups. As we argue in more detail later, this assumption might not be applicable in this context. The bounds we propose do not rely on this assumption, but they are wide and uninformative, here again because schooling increased in the author's control group. Therefore, we form a new control group where years of schooling did not change. The Wald-DID with our new groups is twice as large as the author's original estimate. The Wald-TC and Wald-CIC lie in-between the two. The Wald-DID relies on the assumption that returns to schooling do not change over cohorts, which rules out decreasing returns to experience. Because the Wald-TC and Wald-CIC do not rely on this assumption, we choose them as our favorite estimates.

\medskip
Overall, our paper shows that researchers who use the DID method with fuzzy groups can obtain estimates not resting on any treatment effect homogeneity assumption, provided they can find a control group whose exposure to the treatment does not change over time.

\medskip
Though  we are the first to study fuzzy DID estimators in models with heterogeneous treatment effects, our paper is related to several other papers in the DID  literature. \cite{Blundell04} and \cite{Abadie05} consider a conditional version of the common trends assumption in sharp DID designs, and adjust for covariates using propensity score methods. Our Wald-DID estimator with covariates is related to their estimators. \cite{bonhomme2011} consider a linear model allowing for heterogeneous effects of time, and
show that in sharp designs it can be identified if the idiosyncratic shocks are independent of the treatment and of the individual effects. Our Wald-CIC estimator builds on \cite{Athey06}. In work posterior to ours, \cite{Dhault13} study the possibly nonlinear effects of a continuous treatment, and propose an estimator related to our Wald-CIC estimator.

\medskip
The remainder of the paper is organized as follows. Section 2 presents our main identification results in a simple setting with two groups, two periods, and a binary treatment. Section 3 presents identification results in applications with more than two groups, as well as other extensions. Section 4 presents estimation and inference. In section 5 we revisit results from \cite{Duflo01}. Section 6 concludes. The appendix gathers the main proofs. Due to a concern for brevity, further identification and inference results, two additional empirical applications, and additional proofs are deferred to our supplementary material.

\section{Identification}
\label{sec:ident}

\subsection{Framework}
\label{sub:framework}

We are interested in measuring the effect of a treatment $D$ on some outcome. For now, we assume that the treatment is binary. $Y(1)$ and $Y(0)$  denote the two potential outcomes of the same unit with and without treatment. The observed outcome is $Y=D Y(1) +(1-D)Y(0)$.

\medskip
Hereafter, we consider a model best suited for repeated cross sections. This model also applies to single cross sections where cohort of birth plays the role of time, as in \cite{Duflo01} for instance. The extension to panel data is sketched in Subsection \ref{sec:others} and developed in our supplementary material. We assume that the data can be divided into ``time periods'' represented by a random variable $T$, and into groups represented by a random variable $G$. In this section and in the next, we focus on the simplest possible case where there are only two groups, a ``treatment'' and a ``control'' group, and two periods of time. $G$  is a dummy for units in the treatment group and $T$ is a dummy for the second period.

\medskip
We now introduce the notation we use throughout the paper. For any random variable $R$, let $\Supp(R)$ denote its support. Let also $R_{gt}$ and $R_{dgt}$ be two other random variables such that $R_{gt} \sim R |\,G=g, T=t$ and $R_{dgt} \sim R |\,D=d, G=g, T=t$, where $\sim$ denotes equality in distribution. For instance, it follows from those definitions that $E(R_{11})=E(R|G=1,T=1)$, while $E(R_{011})=E(R|D=0,G=1,T=1)$. For any event or random variable $A$, let $F_{R}$ and $F_{R|A}$ denote the cumulative distribution function (cdf) of $R$ and its cdf conditional on $A$.\footnote{With a slight abuse of notation, $P(A) F_{R|A}$ should be understood as $0$ when $A$ is an event and $P(A)=0$.} Finally, for any increasing function $F$ on the real line, we denote by $F^{-1}$ its generalized inverse, $F^{-1}(q)=\inf\left\{ x\in\R:F(x)\geq q\right\}$. In particular, $F_R^{-1}$ is the quantile function of the random variable $R$.

\medskip
Contrary to the standard ``sharp'' DID setting where $D= G \times T$,  we consider a ``fuzzy'' setting where $D\neq G \times T$. Some units may be treated in the control group or at period $0$, and some units may remain untreated in the treatment group at period $1$. However, we assume that the treatment rate increased more between period 0 and 1 in the treatment than in the control group. Accordingly, we introduce two assumptions we maintain throughout the paper.

\begin{hyp} \label{hyp:timeinvariance2}
(Treatment participation equation)
	
$D = 1\{V\geq v_{GT}\}$, with $V \indep T|G$.
\end{hyp}

\begin{hyp}\label{hyp:first_stage}
(First stage)

$E(D_{11})>E(D_{10})$, and $E(D_{11})-E(D_{10})>E(D_{01})-E(D_{00})$.
\end{hyp}

Assumption \ref{hyp:timeinvariance2} imposes a latent index model for the treatment (see, e.g., \citep{Vytlacil02}), where the threshold depends both on time and group.\footnote{This selection equation implies that within each group, units can switch treatment in only one direction. While this assumption greatly simplifies the presentation of our identification results, it is not necessary for them to hold. See our discussion of panel data in Subsection \ref{sec:others} for further detail on this point.} $V$ may be interpreted as a unit's propensity to be treated. Assumption \ref{hyp:timeinvariance2} also imposes that the distribution of $V$ is stable within each group. Assumption \ref{hyp:first_stage} is just a way to define the treatment and the control group in our fuzzy setting. The treatment group is the one experiencing the larger increase of its treatment rate. If the treatment rate decreases in both groups, one can redefine the treatment variable as $\tilde{D}= 1-D$. Thus, Assumption \ref{hyp:first_stage} only rules out the case where the two groups experience the same evolution of their treatment rates.

\medskip
We now define our parameters of interest. For that purpose, let us introduce
$$D(t) = 1\{V\geq v_{Gt}\}.$$
In repeated cross sections, $D(0)$ and $D(1)$ denote the treatment status of a unit at period 0 and 1 respectively, and only $D=D(T)$ is observed. In single cross sections where cohort of birth plays the role of time, $D(t)$ denotes instead the potential treatment of a unit had she been born at $T=t$. Here again, only $D=D(T)$ is observed. Then, let $S=\{D(0) < D(1),G=1\}$. $S$ stands for treatment group units going from non treatment to treatment between period 0 and 1, hereafter referred to as the ``treatment group switchers''. Our parameters of interest are their Local Average Treatment Effect (LATE) and Local Quantile Treatment Effects (LQTE), which are respectively defined by
\begin{eqnarray*}
\Delta & = & E\left(Y_{11}(1) - Y_{11}(0)|S\right), \\
\tau_q & =& F_{Y_{11}(1)|S}^{-1}(q) - F_{Y_{11}(0)|S}^{-1}(q), \quad q \in (0,1).
\end{eqnarray*}
We focus on these parameters for two reasons. First, there are instances where treatment group switchers are the only units affected by some policy, implying that they are the relevant subgroup one should consider to assess its effects. Consider for instance a policy whereby in $T=1$, the treatment group becomes eligible to some treatment for which it was not eligible in $T=0$ (see, e.g., \citep{field2007}). In this example, treatment group switchers are all the units in that group treated in $T=1$. Those units are affected by the policy: without it, they would have remained untreated. Moreover, nobody else is affected by the policy. Second, identifying treatment effects in the whole population would require additional conditions, on top of those we consider below. In the example above, the policy extension does not provide any information on treatment effects in the control group, because this group does not experience any change.

\subsection{The Wald differences-in-differences estimand}\label{sec:IVDID}

We first investigate the commonly used strategy of running an IV regression of the outcome on the treatment with time and group as included instruments, and the interaction of the two as the excluded instrument. The estimand arising from this regression is the Wald-DID defined by $W_{DID}=DID_Y/DID_D,$ where for any random variable $R$ we let
$$DID_R=E(R_{11})-E(R_{10})-\left(E(R_{01})-E(R_{00})\right).$$
Let also $S'=\{D(0) \neq D(1), G=0\}$ denote the control group switchers, and let $\Delta'=E\left(Y_{01}(1)-Y_{01}(0)|S'\right)$ denote their LATE. Finally, let $\alpha=\left(P(D_{11}=1)-P(D_{10}=1)\right)/DID_D$.

\medskip
We consider the following assumptions, under which we can relate $W_{DID}$ to $\Delta$ and $\Delta'$.
\begin{hyp}\label{hyp:common_trends_between}
(Common trends)

\medskip
$E(Y(0)|G,T=1)-E(Y(0)|G,T=0)$ does not depend on $G$.
\end{hyp}
\begin{hyp}\label{hyp:common_trends_within}
(Homogeneous treatment effect over time)

\medskip
For all $d\in \Supp(D)$,\footnote{When the treatment is binary, Assumption \ref{hyp:common_trends_within} only requires that the equation therein holds for $d=1$. Writing Assumption \ref{hyp:common_trends_within} this way ensures it carries through to the case of a non-binary treatment.} $E(Y(d)-Y(0)|G, T=1, D(0)=d)=E(Y(d)-Y(0)|G, T=0, D(0)=d).$
\end{hyp}
\begin{hyp}\label{hyp:hte}
(Homogeneous treatment effect between groups)

\medskip
$\Delta=\Delta'.$
\end{hyp}
\begin{hyp}
(Stable percentage of treated units in the control group)

$0<E(D_{01})=E(D_{00})<1$.
\label{hyp:stable_control}
\end{hyp}
Assumption \ref{hyp:common_trends_between} requires that the mean of $Y(0)$ follow the same evolution over time in the treatment and control groups.
This assumption is not specific to the fuzzy settings we are considering here: DID in sharp settings
also rely on this assumption (see, e.g., \citep{Abadie05}). Assumption \ref{hyp:common_trends_within}
requires that in both groups, the average treatment effect among
units treated in period $0$ is stable over time. This is equivalent to
assuming that among these units, the mean of $Y(1)$ and $Y(0)$ follow the same evolution over time:
\begin{eqnarray*}
&& E(Y(1)|G,T=1, D(0)=1)-E(Y(1)|G,T=0, D(0)=1) \\
&=& E(Y(0)|G,T=1, D(0)=1)-E(Y(0)|G,T=0, D(0)=1).	
\end{eqnarray*}
This assumption is specific to fuzzy settings. Assumption \ref{hyp:hte} requires that in both groups, switchers have the same LATE. This assumption is also specific to fuzzy settings. Finally, Assumption \ref{hyp:stable_control} requires that the share of treated units in the control group does not change between period 0 and 1 and is included between 0 and 1. While Assumptions \ref{hyp:common_trends_between} to \ref{hyp:hte} are not directly testable, Assumption \ref{hyp:stable_control} can be assessed from the data.
\begin{thm}\label{thm:IV-DID1} ~
\begin{enumerate}
\item If Assumptions \ref{hyp:timeinvariance2}-\ref{hyp:common_trends_within} are satisfied, then
\begin{align*}
W_{DID} = & \alpha \Delta+(1-\alpha) \Delta'.
\end{align*}
\item If Assumption \ref{hyp:hte} or  \ref{hyp:stable_control} further holds, then
\begin{align*}
W_{DID} = & \Delta.
\end{align*}
\end{enumerate}
\end{thm}
When the treatment rate increases in the control group, $E(D_{01})-E(D_{00})>0$, so $\alpha>1$. Therefore, under Assumptions \ref{hyp:timeinvariance2}-\ref{hyp:common_trends_within} the Wald-DID is
equal to a weighted difference of the LATEs of treatment and control group switchers in
period 1. In both groups, the evolution of the mean outcome between period 0 and 1 is the sum
of three things: the change in the mean of $Y(0)$ for units untreated at $T=0$; the change in
the mean of $Y(1)$ for units treated at $T=0$; the average effect of the treatment for switchers. Under
Assumptions \ref{hyp:common_trends_between} and \ref{hyp:common_trends_within}, changes in the mean
of $Y(0)$ and $Y(1)$ in both groups cancel out. The Wald-DID is finally equal to the weighted
difference between the LATEs of treatment and control group switchers. This weighted difference
does not satisfy the no sign-reversal property: it may be negative even if the treatment effect is
positive for everybody in the population. If one is ready to further assume that Assumption
\ref{hyp:hte} is satisfied, this weighted difference simplifies into $\Delta$.\footnote{Under
this assumption, the Wald-DID actually identifies the LATE of all switchers, not only of those in the treatment group.
There are instances where this LATE measures the effect of the policy under consideration, because treatment and control group switchers are the only units affected by this policy. Consider for instance the case of a policy whereby a new treatment is introduced in both groups in $T=1$ (see \citep{enikolopov2011}). In this example, treatment and control group switchers are all the units treated in $T=1$. These units are affected by the policy (without it, they would have remained untreated) and nobody else is affected by it.}

\medskip
When the treatment rate diminishes in the control group, $E(D_{01})-E(D_{00})<0$, so $\alpha<1$. Therefore, under Assumptions \ref{hyp:timeinvariance2}-\ref{hyp:common_trends_within} the Wald-DID is equal to a
weighted average of the LATEs of treatment and control group switchers in period 1. This quantity
satisfies the no sign-reversal property, but it still differs from $\Delta$ unless here as well one is ready to further assume that Assumption \ref{hyp:hte} is satisfied.

\medskip
When the treatment rate is stable in the control group, $\alpha=1$ so the Wald-DID is equal to $\Delta$ under Assumptions \ref{hyp:timeinvariance2}-\ref{hyp:common_trends_within} alone.
But even then, the Wald-DID relies on a treatment effect homogeneity assumption: in both groups, the average treatment effect among units treated at $T=0$ should remain stable over time. This assumption is necessary. Under Assumptions \ref{hyp:timeinvariance2}-\ref{hyp:common_trends_between} alone, the Wald-DID is equal to $\Delta$ plus a bias term involving several LATEs. Unless this combination of LATEs cancels out exactly, the Wald-DID differs from $\Delta$. We give the formula of the bias term in the end of the proof of Theorem \ref{thm:IV-DID1}.

\subsection{The time-corrected Wald estimand}\label{sub:Wald_TC}

In this section, we consider a first alternative estimand of $\Delta$.
Instead of relying on Assumptions \ref{hyp:common_trends_between} and \ref{hyp:common_trends_within}, it is based on the following condition:

\setcounter{hyp}{2}

\begin{hyp}\hspace{-0.2cm}{}' \label{hyp:common_trends_conditional}
(Conditional common trends)

\medskip
For all $d\in \Supp(D)$, $E(Y(d)|G,T=1,D(0)=d)- E(Y(d)|G,T=0,D(0)=d)$ does not depend on $G$.
\end{hyp}

\setcounter{hyp}{6}

Assumption \ref{hyp:common_trends_conditional}' requires that the mean of $Y(0)$ (resp. $Y(1)$) follows the same evolution over time among treatment and control group units that were untreated (resp. treated) at $T=0$.

\medskip
Let $\delta_d=E(Y_{d01})-E(Y_{d00})$ denote the change in the mean outcome between period 0 and 1 for control group units with treatment status $d$. Then, let
\begin{eqnarray*}
&&W_{TC}=\frac{E(Y_{11})-E(Y_{10}+\delta_{D_{10}})}{E(D_{11})-E(D_{10})}.
\end{eqnarray*}
$W_{TC}$ stands for ``time-corrected Wald''.

\begin{thm}\label{thm:IV-DID2}
If Assumptions \ref{hyp:timeinvariance2}-\ref{hyp:first_stage}, \ref{hyp:common_trends_conditional}', and \ref{hyp:stable_control} are satisfied, then $W_{TC}=\Delta.$
\end{thm}

Note that $$W_{TC}=\frac{E(Y|G=1,T=1)-E(Y+(1-D)\delta_0+D\delta_1|G=1,T=0)}{E(D|G=1,T=1)-E(D|G=1,T=0)}.$$
This is almost the Wald ratio in the treatment group with time as the instrument, except that we have $Y+(1-D)\delta_0+D\delta_1$ instead of $Y$ in the second term of the numerator. This difference arises because time is not a standard instrument: it can directly affect the outcome. When the treatment rate is stable in the control group, we can identify the trends on $Y(0)$ and $Y(1)$ by looking at how the mean outcome of untreated and treated units changes over time in this group. Under Assumption \ref{hyp:common_trends_conditional}', these trends are the same in the two groups. As a result, we can add these changes to the outcome of untreated and treated units in the treatment group in period $0$, to recover the mean outcome we would have observed in this group in period $1$ if switchers had not changed their treatment between the two periods. This is what $(1-D)\delta_0+D\delta_1$ does. Therefore, the numerator of $W_{TC}$ compares the mean outcome in the treatment group in period 1 to the counterfactual mean we would have observed if switchers had remained untreated. Once normalized, this yields the LATE of treatment group switchers.

\subsection{The changes-in-changes estimands}\label{sec:IVCIC}

In this section, we consider a second alternative estimand of $\Delta$ for continuous outcomes, as well as estimands of the LQTE. They rely on the following condition.

\begin{hyp}\label{hyp:monotonicity1}
(Monotonicity and time invariance of unobservables)

\medskip
$Y(d)=h_d(U_d,T)$, with $U_d\in \R$ and $h_d(u,t)$ strictly increasing in $u$ for all $(d,t) \in \Supp((D,T))$. Moreover, $U_d \indep T | G,D(0)$.
\end{hyp}

Assumptions \ref{hyp:timeinvariance2} and \ref{hyp:monotonicity1} generalize the CIC model in \cite{Athey06} to fuzzy settings.
Assumptions \ref{hyp:timeinvariance2} and \ref{hyp:monotonicity1} imply $U_{d}\indep T\,|\, G$. Therefore, they require that at each period, both potential outcomes are strictly increasing functions of a scalar unobserved heterogeneity term whose distribution is stationary over time, as in \cite{Athey06}. But Assumption \ref{hyp:monotonicity1} also imposes $U_{d}\indep T\,|\, G,D(0)$: the distribution of $U_{d}$ must be stationary within subgroup of units sharing the same treatment status at $T=0$.

\medskip
We also impose the assumption below, which is testable in the data.
\begin{hyp}\label{hyp:data1}
(Data restrictions)
\begin{enumerate}
\item $\Supp(Y_{dgt}) =\Supp(Y)$ for
$(d,g,t)\in \Supp((D,G,T))$, and $\Supp(Y)$ is a closed interval of $\mathbb{R}$.
\item $F_{Y_{dgt}}$ is continuous on $\mathbb{R}$ and strictly increasing on $\Supp(Y)$, for
$(d,g,t)\in \Supp((D,G,T))$.
\end{enumerate}
\end{hyp}
The first condition requires that the outcome have the same support in each of the eight treatment $\times$ group $\times$ period cells. \cite{Athey06} make a
similar assumption.\footnote{Common support conditions might not be satisfied when outcome distributions differ in the treatment and control groups, the very situations where the Wald-CIC estimand we propose below might be more appealing than the Wald-DID or Wald-TC (see Subsection \ref{sec:comparison}). \cite{Athey06} show that in such instances, quantile treatment effects are still point identified over a large set of quantiles, while the average treatment effect can be bounded. Even though we do not present them here, similar results apply in fuzzy settings.} Note that this condition does not restrict the outcome to have bounded support: for instance, $[0,+\infty)$ is a closed interval of $\mathbb{R}$. The second condition requires that the distribution of $Y$ be continuous with positive density in each of the eight groups $\times$ periods $\times$ treatment status cells. With a discrete outcome, \cite{Athey06} show that one can bound treatment effects under their assumptions. Similar results apply in fuzzy settings, but for the sake of brevity we do not present them here.

\medskip
Let $Q_{d}(y)=F^{-1}_{Y_{d01}} \circ F_{Y_{d00}}(y)$ be the quantile-quantile transform of $Y$
from period 0 to 1 in the control group conditional on $D=d$. This transform maps $y$ at rank
$q$ in period $0$ into the corresponding $y'$ at rank $q$ in period 1. Let also
\begin{align*}
F_{CIC,d}(y)& =\frac{P(D_{11}=d)F_{Y_{d11}}(y) - P(D_{10}=d) F_{Q_d(Y_{d10})}(y)}{P(D_{11}=d)-P(D_{10}=d)},\\
W_{CIC}& =\frac{E(Y_{11}) - E(Q_{D_{10}}(Y_{10}))}{E(D_{11}) - E(D_{10})}.
\end{align*}

\begin{thm}
\label{thm:ident_CIC} If Assumptions \ref{hyp:timeinvariance2}-\ref{hyp:first_stage}, \ref{hyp:stable_control}, and \ref{hyp:monotonicity1}-\ref{hyp:data1} are satisfied, then $W_{CIC}=\Delta$ and $F^{-1}_{CIC,1}(q)-F^{-1}_{CIC,0}(q)=\tau_q$.
\end{thm}

This result combines ideas from \cite{Imbens97} and \cite{Athey06}. We seek to recover the distribution of, say, $Y(1)$ among switchers in the treatment group $\times$ period 1 cell. On that purpose, we start from the distribution of $Y$ among all treated observations of this cell. Those include both switchers and units already treated at $T=0$. Consequently, we must ``withdraw'' from this distribution that of units treated at $T=0$, exactly as in \cite{Imbens97}. But this last distribution is not observed.  To reconstruct it, we adapt the ideas in \cite{Athey06} and apply the quantile-quantile transform from period 0 to 1 among treated observations in the control group to the distribution of $Y$ among treated units in the treatment group in period 0.

\medskip
Intuitively, the quantile-quantile transform uses a double-matching to reconstruct the unobserved distribution. Consider a treated unit in the treatment group $\times$ period 0 cell. She is first matched to a treated unit in the control group $\times$ period 0 cell with same $y$. Those two units are observed at the same period of time and are both treated. Therefore, under Assumption \ref{hyp:monotonicity1}  they must have the same $u_1$. Second, the control $\times$ period $0$ unit is matched to her rank counterpart among treated units of the control group $\times$ period 1 cell. We denote by $y^{*}$ the outcome of this last observation. Because $U _1 \indep T |G, D(0)=1$, under Assumption \ref{hyp:stable_control} those two observations must also have the same $u_1$. Consequently, $y^{*}=h_1(u_1,1)$, which means that $y^{*}$ is the outcome that the treatment $\times$ period 0 cell unit would have obtained in period 1.

\medskip
Note that $$W_{CIC}=\frac{E(Y|G=1,T=1) - E((1-D)Q_0(Y)+DQ_1(Y)|G=1,T=0)}{E(D|G=1,T=1) - E(D|G=1,T=0)}.$$
Here again, $W_{CIC}$ is almost the standard Wald ratio in the treatment group with $T$ as the instrument, except that we have $(1-D)Q_0(Y)+DQ_1(Y)$ instead of $Y$ in the second term of the numerator. $(1-D)Q_0(Y)+DQ_1(Y)$ accounts for the fact that time directly affects the outcome, just as $(1-D)\delta_0+D\delta_1$ does in the $W_{TC}$ estimand. Under Assumption \ref{hyp:common_trends_conditional}', the trends affecting the outcome are identified by additive shifts, while under Assumptions \ref{hyp:monotonicity1}-\ref{hyp:data1} they are identified by possibly non-linear quantile-quantile transforms.

\subsection{Choosing between the Wald-DID, Wald-TC, and Wald-CIC estimands}\label{sec:comparison}

When the treatment rate is stable in their control group, researchers need to chose between the three estimands we have proposed in this section. In order to do so, they can start by conducting placebo tests. Assume for instance that data is available for period $T=-1$, and that the share of treated units is stable in both groups between $T=-1$ and $T=0$:
$E(D_{10})-E(D_{1-1})=E(D_{00})-E(D_{0-1})=0$. Then Assumptions \ref{hyp:common_trends_between} and
\ref{hyp:common_trends_within} between $T=-1$ and 0 imply that $E(Y_{10})-E(Y_{1-1})-(E(Y_{00})-E(Y_{0-1}))=0$.
Similarly, Assumption \ref{hyp:common_trends_conditional}' (resp. Assumption \ref{hyp:monotonicity1}) between
$T=-1$ and 0 implies that $E(Y_{10})-E(Y_{1-1}+\delta_{D_{1-1}})=0$ (resp. $E(Y_{10})-E(Q_{D_{1-1}}(Y_{1-1}))
=0$).\footnote{With a slight abuse of notation, here $\delta_d$ and $Q_d$ are computed between periods -1 and
0.} Assumptions \ref{hyp:common_trends_conditional}' and \ref{hyp:monotonicity1} have further testable implications. For instance, Assumption \ref{hyp:common_trends_conditional}' implies that for $d \in \{0,1\}$, $E(Y_{10d})-E(Y_{1-1d})=E(Y_{00d})-E(Y_{0-1d})$: common trends between the two groups should hold conditional on each value of the treatment.

\medskip
On the other hand, when $E(D_{10})-E(D_{1-1})$ or $E(D_{00})-E(D_{0-1})$ is different from 0, placebo
estimators can no longer be used to test Assumptions \ref{hyp:common_trends_between} and
\ref{hyp:common_trends_within}, \ref{hyp:common_trends_conditional}', or \ref{hyp:monotonicity1}. Placebos
might differ from zero even if those assumptions are satisfied, because of the effect of the treatment.

\medskip
Sometimes, even after using placebos to discard estimators relying on implausible assumptions, researchers
might be left with several, significantly different estimators. This can be due to lack of power. This can
also be due to the fact that placebos are tests of Assumptions \ref{hyp:common_trends_between}-\ref{hyp:common_trends_within}, \ref{hyp:common_trends_conditional}', or \ref{hyp:monotonicity1} for pairs
of dates prior to $T=1$, while our estimands require that these assumptions hold between $T=0$ and 1.

\medskip
In such instances, inspecting the assumptions underlying each estimand through the lens of economic theory
might help researchers choose between the Wald-DID and Wald-TC estimands. In applications where technological or institutional evolutions make it likely that treatment effects change over time, it might be appealing to choose the Wald-TC estimand, so as not to rely on Assumption \ref{hyp:common_trends_within}.  On the other hand, Assumption \ref{hyp:common_trends_conditional}' may be more restrictive than Assumption \ref{hyp:common_trends_between}, in particular when units self-select themselves into treatment. One might for instance worry that the treatment rate increases in the treatment group because units in this
group experience a positive shock on their $Y(1)$ at $T=1$. This would imply that Assumption
\ref{hyp:common_trends_conditional}' is violated while Assumption \ref{hyp:common_trends_between} might hold.
But in this scenario, Assumption \ref{hyp:common_trends_within} would also be violated, thus implying
that both the Wald-DID and the Wald-TC estimators are inconsistent.\footnote{Note however that Assumptions
\ref{hyp:common_trends_within} and \ref{hyp:common_trends_conditional}' are not incompatible with Roy models of selection into treatment. For
instance, if $D=1\{Y(1)-Y(0)\geq c(G,T)\}$, $Y(d)= a_{dG}+bT+e_d$, $(e_0,e_1)\indep T|G$, and $E(e_d|G)=0$,
then Assumptions \ref{hyp:common_trends_within} and \ref{hyp:common_trends_conditional}' are satisfied. On the other hand, if  $Y(d)= a_{dG}+bT
+\sigma_Te_d$ with $\sigma_0\ne \sigma_1$, then Assumptions \ref{hyp:common_trends_within} and \ref{hyp:common_trends_conditional}' fail.}

\medskip
On the other hand, the assumptions underlying the Wald-CIC and Wald-TC estimands are substantively close.\footnote{For this reason, one can follow the same steps as those outlined in the previous paragraph to choose between the Wald-DID and Wald-CIC estimands.} Therefore, economic theory can provide little guidance as to which estimand one should pick. Here the choice should rather be based on whether the treatment and the control groups have very different outcome distributions conditional on $D$ at $T=0$.
Assumption \ref{hyp:common_trends_conditional}' is not invariant to the scaling of the outcome, but it only restricts its first moment. Assumption \ref{hyp:monotonicity1}
is invariant to the scaling of the outcome, but it restricts its entire distribution. When the treatment and the control groups have different outcome distributions conditional on $D$ in the first period (see e.g. \citep{baten2014}), the scaling of the outcome might have a large effect on the Wald-TC, so using the Wald-CIC might be preferable. On the other hand, when the two groups have similar outcome distributions conditional on $D$ in the first period, using the Wald-TC might be preferable as it only restricts the first moment of the outcome. Another advantage of working under Assumptions \ref{hyp:timeinvariance2} and \ref{hyp:monotonicity1} is that this enables the analyst to study distributional effects instead of mean effects only.

\medskip
Finally, when the treatment rate varies in the control group, the assumptions underlying the Wald-TC and Wald-CIC estimands only lead to partial identification (see Subsection \ref{sec:partial}), and the bounds may not be informative, as is the case in our application below. On the other hand, the Wald-DID estimand can still point identify $\Delta$. This estimand may then be appealing, especially when the treatment rate decreases in the control group, as in such instances it estimates a weighted average of LATEs even if Assumption \ref{hyp:hte} fails to hold. When the treatment rate increases in the control group, Assumption \ref{hyp:hte} is necessary to have $W_{DID}=\Delta$, and placebo tests are generally uninformative as to the plausibility of this assumption. To see this, assume for instance that a treatment appears in $T=1$ and that some units are treated both in the treatment and in the control group. This corresponds to the situation in \cite{enikolopov2011}, who study the effect of the introduction of an independent TV channel in Russia on votes for the opposition. In such instances, placebo DIDs comparing the evolution of the mean outcome in the two groups before $T=1$ are tests of Assumption \ref{hyp:common_trends_between}, but they are uninformative as to the plausibility of Assumption \ref{hyp:hte} because nobody was treated before $T=1$. Therefore, this assumption should be carefully discussed.

\section{Extensions}
\label{sec:extension}

We now consider several extensions. We first consider applications with multiple groups. We then show that our results extend to ordered, non-binary treatments. Next, we show that when the treatment rate is not stable in the control group, $\Delta$ and $\tau_q$ can still be partially identified under our assumptions. Finally we sketch other extensions that are fully developed in the supplementary material.

\subsection{Multiple groups}\label{sec:multiple_time_periods}

We consider the case where there are more than two groups but only two time periods in the data. The case
with multiple groups and time periods is considered in the supplementary material. Let
$G\in \{0,1,...,\overline{g}\}$ denote the group a unit belongs to.
For any $g\in \Supp(G)$, let $S_g=\{D(0)\neq D(1),G=g\}$ denote units of group $g$ who switch treatment between $T=0$ and 1. Let  $S^*=\cup_{g=0}^{\overline{g}} S_g$ be the union of all switchers.
We can partition the groups depending on whether their treatment rate is stable, increases, or decreases. Specifically, let
\begin{eqnarray*}
&&\mathcal{G}_{s}=\{g\in \Supp(G): E(D_{g1})=E(D_{g0})\}\\
&&\mathcal{G}_{i}=\{g\in \Supp(G): E(D_{g1})>E(D_{g0})\}\\
&&\mathcal{G}_{d}=\{g\in \Supp(G): E(D_{g1})<E(D_{g0})\},
\end{eqnarray*}
and let $G^*=1\{G\in \mathcal{G}_{i}\}-1\{G\in \mathcal{G}_{d}\}$.

\medskip
Theorem \ref{thm:manygroups} below shows that when there is at least
one group in which the treatment rate is stable, our
assumptions allow us to point identify $\Delta^*=E(Y(1)-Y(0)|S^*,T=1)$, the LATE of all switchers.
Before presenting this result, additional notation is needed. For any random variable $R$,
$g\ne g'\in \{-1,0,1\}^2$, and $d \in \{0,1\}$, let
\begin{eqnarray*}
DID^*_R(g,g')&=&E(R|G^*=g,T=1)-E(R|G^*=g,T=0)\\
&-&(E(R|G^*=g',T=1)-E(R|G^*=g',T=0)),\\
\delta^*_{d}& =& E(Y|D=d,G^*=0,T=1)-E(Y|D=d,G^*=0,T=0), \\
Q^*_{d}(y)& =& F^{-1}_{Y|D=d,G^*=0,T=1} \circ F_{Y|D=d,G^*=0,T=0}(y), \\
W^*_{DID}(g,g')&=&\frac{DID^*_Y(g,g')}{DID^*_D(g,g')},\\
W^*_{TC}(g) & =& \frac{E(Y|G^*=g,T=1)-E(Y+\delta^*_{D}|G^*=g,T=0)}{E(D|G^*=g,T=1)-E(D|G^*=g,T=0)},\\
W^*_{CIC}(g) & =& \frac{E(Y|G^*=g,T=1)-E(Q^*_{D}(Y)|G^*=g,T=0)}{E(D|G^*=g,T=1)-E(D|G^*=g,T=0)}.
\end{eqnarray*}

We also define the following weight:
\begin{eqnarray*}
w_{10}&=&\frac{DID^*_D(1,0)P(G^*=1)}{DID^*_D(1,0)P(G^*=1)+DID^*_D(0,-1)P(G^*=-1)}.
\end{eqnarray*}

\begin{thm}\label{thm:manygroups}
Assume that Assumption \ref{hyp:timeinvariance2}	 is satisfied, that
$\mathcal{G}_{s} \ne \emptyset$, and that $G\indep T$.
\begin{enumerate}
\item If Assumptions \ref{hyp:common_trends_between} and \ref{hyp:common_trends_within} are satisfied,
\begin{align*}
w_{10}W^*_{DID}(1,0)+(1-w_{10})W^*_{DID}(-1,0) & = \Delta^*.
\end{align*}
\item If Assumption \ref{hyp:common_trends_conditional}' is satisfied,
\begin{align*}
w_{10}W^*_{TC}(1)+(1-w_{10})W^*_{TC}(-1) = & \Delta^*.
\end{align*}
\item If Assumptions \ref{hyp:monotonicity1} and \ref{hyp:data1} are satisfied,
\begin{align*}
	w_{10}W^*_{CIC}(1)+(1-w_{10})W^*_{CIC}(-1) = & \Delta^*.
\end{align*}
\end{enumerate}
\end{thm}
This theorem states that with multiple groups and two periods of time, treatment effects for
switchers are identified if there is at least one group in which the treatment rate is stable
over time. The estimands we propose can then be
computed in four steps. First, we form three ``supergroups'', by pooling together the groups
where treatment increases ($G^*=1$), those where it is stable ($G^*=0$), and those where it
decreases ($G^*=-1$). While in some applications these three sets of groups are known to the
analyst (see e.g. \citep{gentzkow2011}), in other applications they must be estimated (see our application in Section \ref{sec:applications2}).
Second, we compute the Wald-DID, Wald-TC, or Wald-CIC estimand with $G^*=1$ and $G^*=0$ as the treatment and control groups. Third, we
compute the Wald-DID, Wald-TC, or Wald-CIC estimand with $G^*=-1$ and $G^*=0$ as the
treatment and control groups. Finally, we compute a weighted average of those two estimands.

\medskip
Theorem \ref{thm:manygroups} relies on the Assumption that $G\indep T$. This requires that the distribution of
groups be stable over time. This will automatically be satisfied if the data is a balanced panel and $G$ is time
invariant. With repeated cross-sections or cohort data, this assumption might fail to hold. However,
when $G$ is not independent of $T$, it is still possible to form Wald-DID and Wald-TC type of estimands identifying $\Delta^*$. We give the formulas of these estimands in Subsection \ref{sub:multiple_periods} in the supplementary material.

\medskip
Three last comments on Theorem \ref{thm:manygroups} are in order. First, it contrasts with the current practice in empirical work. When many groups are available, researchers usually include group fixed effects in their regressions, instead of pooling together groups into super control and treatment groups as we advocate here. In \cite{deChaisemartin16}, we show that such regressions estimate a weighted sum of switchers' LATEs across groups, with potentially many negative weights and without the aggregation property we obtain here. Second, groups where the treatment rate diminishes can be used as ``treatment'' groups, just as those where it increases. Indeed, it is easy to show that all the results from the previous section still hold if the treatment rate decreases in the treatment group and is stable in the control group. Finally, when there are more than two groups where the treatment rate is stable, our three sets of assumptions become testable. Under each set of assumptions, using any subset of $\mathcal{G}_{s}$ as the control group should yield the same estimand for $\Delta^*$.

\subsection{Non-binary, ordered treatment}\label{sec:multivariate_treatment}

We now consider the case where treatment takes a finite number of ordered values, $D \in \{0,1,...,\overline{d}\}$. To accommodate this extension, Assumption \ref{hyp:timeinvariance2} has to be modified as follows.

\setcounter{hyp}{0}

\begin{hyp} \label{hyp:ordered} \hspace{-0.2cm}{}' \;
(Ordered treatment equation)

\medskip
$D = \sum_{d=1}^{\overline{d}}1\{V\geq v^{d}_{GT}\}$, with
$-\infty = v^0_{GT} < v^1_{GT}...<v^{\overline{d}+1}_{GT}=+\infty$ and $V \indep T|G$. As before, let $D(t)=\sum_{d=1}^{\overline{d}}1\{V\geq v^{d}_{Gt}\}$.
\end{hyp}

\setcounter{hyp}{8}

Let $\gtrsim$ denote stochastic dominance between two random variables, and let $\sim$ denote equality in distribution. Let also $w_d=[P(D_{11}\geq d)-P(D_{10}\geq d)]/[E(D_{11})-E(D_{10})]$.

\begin{thm}\label{thm:multivariate}
Suppose that Assumption \ref{hyp:ordered}' and \ref{hyp:first_stage} hold, that $D_{01}\sim D_{00}$, and that
$D_{11}\gtrsim D_{10}$.
\begin{enumerate}
\item If Assumptions \ref{hyp:common_trends_between}-\ref{hyp:common_trends_within}  are satisfied,
\begin{align*}
W_{DID} = & \sum_{d=1}^{\overline{d}}w_d E(Y_{11}(d)-Y_{11}(d-1)|D(0) < d \leq D(1)).
\end{align*}
\item If Assumption \ref{hyp:common_trends_conditional}' is satisfied,
\begin{align*}
W_{TC} = & \sum_{d=1}^{\overline{d}}w_dE(Y_{11}(d)-Y_{11}(d-1)|D(0) < d \leq D(1)).
\end{align*}
\item If Assumptions \ref{hyp:monotonicity1} and \ref{hyp:data1} are satisfied,
\begin{align*}
W_{CIC} = & \sum_{d=1}^{\overline{d}}w_dE(Y_{11}(d)-Y_{11}(d-1)|D(0) < d \leq D(1)).
\end{align*}
\end{enumerate}
\end{thm}

Theorem \ref{thm:multivariate} shows that with an ordered treatment, the estimands we considered
in the previous sections are equal to the average causal response (ACR) parameter considered in
\cite{Angrist95}. This parameter is a weighted average, over all values of $d$, of the effect of
increasing treatment from $d-1$ to $d$ among switchers whose treatment status goes from strictly
below to above $d$ over time.

\medskip
For this theorem to hold, two conditions have to be satisfied. First, in the treatment group, the
distribution of treatment in period 1 should dominate stochastically the corresponding distribution
in period 0. \cite{Angrist95} impose a similar stochastic dominance condition. Actually, this assumption is not necessary for our three estimands
to identify a weighted sum of treatment effects. If it is not satisfied, one still has that $W_{DID}$,
$W_{TC}$, or $W_{CIC}$ identify
\begin{align*}
& \sum_{d=1}^{\overline{d}}w_dE\left(Y_{11}(d)-Y_{11}(d-1)|D(0) < d \leq D(1) \cup D(1) < d \leq D(0)\right),
\end{align*}
which is a weighted sum of treatment effects with some negative weights.

\medskip
Second, the distribution of treatment should be stable over time in the control group. When it is not,
one can still obtain some identification results. Firstly, Theorem \ref{thm:IV-DID1} generalizes to
non-binary and ordered treatments. When treatment increases in the
control group, the Wald-DID identifies a weighted difference of the ACRs in the treatment and in the
control group; when treatment decreases in the control group, the Wald-DID identifies a weighted
average of these two ACRs. The weights are the same as those in Theorem \ref{thm:IV-DID1}. Secondly,
our partial identification results below also generalize to non-binary and ordered treatments. When the distribution of treatment is not stable
over time in the control group, the ACR in the treatment group can be bounded under Assumption
\ref{hyp:common_trends_conditional}', or Assumptions \ref{hyp:monotonicity1}-\ref{hyp:data1}, as shown in Subsection \ref{sub:bounds_appli} of the supplementary material.

\medskip
Finally, Theorem \ref{thm:multivariate} extends to a continuous treatment. In such instances, one can show that under an appropriate generalization of Assumption \ref{hyp:timeinvariance2}, the Wald-DID, Wald-TC, and Wald-CIC identify a weighted average of the derivative of potential outcomes with respect to treatment, a parameter that resembles that studied in \cite{angrist2000}.

\subsection{Partial identification with a non stable control group}\label{sec:partial}

In this subsection, we come back to our basic set-up with two groups and a binary treatment, and we show that $\Delta$ and $\tau_q$ can still be partially identified when Assumption \ref{hyp:stable_control} does not hold. Let us introduce some additional notation. When the outcome is bounded, let $\underline{y}$ and $\overline{y}$ respectively denote the lower and upper bounds of its support. For any real number $x$, let $M_{01}(x)=\min(1,\max(0,x))$. For any $g \in \Supp(G)$, let $\lambda_{gd}=P(D_{g1}=d)/P(D_{g0}=d)$ be the ratio of the shares of units in group $g$ receiving treatment $d$ in period 1 and period 0. For instance, $\lambda_{00}>1$ when the share of untreated observations increases in the control group between period 0 and 1. Let also
\begin{align*}
&\underline{F}_{d01}(y) = M_{01}\left(1-\lambda_{0d} (1-F_{Y_{d01}}(y))\right) - M_{01}(1-\lambda_{0d})\mathds{1}\{y< \overline{y}\}, \\
&\overline{F}_{d01}(y) = M_{01}\left(\lambda_{0d} F_{Y_{d01}}(y)\right) + (1-M_{01}(\lambda_{0d}))\mathds{1}\{y\geq \underline{y}\}, \\
&\underline{\delta}_d =\int y d\overline{F}_{d01}(y) - E(Y_{d00}), \; \overline{\delta}_d =\int y d\underline{F}_{d01}(y) -E(Y_{d00}),
\end{align*}
We define the bounds obtained under Assumption \ref{hyp:common_trends_conditional}'  (TC bounds hereafter) as follows:
$$\underline{W}_{TC} =\frac{E(Y_{11})-E(Y_{10}+\overline{\delta}_{D_{10}})}{E(D_{11})-E(D_{10})}, \; \overline{W}_{TC}=\frac{E(Y_{11})-E(Y_{10}+\underline{\delta}_{D_{10}})}{E(D_{11})-E(D_{10})}.$$
Next, we define the bounds obtained under Assumptions \ref{hyp:monotonicity1}-\ref{hyp:data1} (CIC bounds hereafter). For $d\in \{0,1\}$ and any cdf $T$, let $H_d=F_{Y_{d10}} \circ F^{-1}_{Y_{d00}}$ and
\begin{align*}
& G_d(T) =  \lambda_{0d} F_{Y_{d01}} + (1-\lambda_{0d})T, \;
C_d(T) = \frac{P(D_{11}=d)F_{Y_{d11}}-P(D_{10}=d)H_d \circ G_d(T)}{P(D_{11}=d)-P(D_{10}=d)}, \\
& \underline{T}_d  = M_{01}\left(\frac{\lambda_{0d} F_{Y_{d01}}-H^{-1}_d(\lambda_{1d}F_{Y_{d11}})}{\lambda_{0d}-1}\right), \; \overline{T}_d = M_{01}\left(\frac{\lambda_{0d} F_{Y_{d01}}-H^{-1}_d(\lambda_{1d}F_{Y_{d11}}+(1-\lambda_{1d}))}{\lambda_{0d}-1}\right),\\
&\underline{F}_{CIC,d}(y) =\sup_{y'\leq y}C_d\left(\underline{T}_d\right)(y'), \; \overline{F}_{CIC,d}(y)=\inf_{y'\geq y}C_d\left(\overline{T}_d\right)(y').
\end{align*}
In the definition of $\underline{T}_d$ and $\overline{T}_d$, we use the convention
that $F_R^{-1}(q) =\inf \Supp(R)$ for $q<0$, and $F_R^{-1}(q) = \sup \Supp(R)$ for $q>1$. We then define the CIC bounds on $\Delta$ and $\tau_q$ by:
\begin{align*}
& \underline{W}_{CIC} = \int y d\overline{F}_{CIC,1}(y) - \int y
d\underline{F}_{CIC,0}(y), \; \overline{W}_{CIC} = \int y d\underline{F}_{CIC,1}(y) - \int y
d\overline{F}_{CIC,0}(y), \\
& \underline{\tau}_q  = \max(\overline{F}_{CIC,1}^{-1}(q), \underline{y}) - \min(\underline{F}_{CIC,0}^{-1}(q), \overline{y}), \;
\overline{\tau}_q = \min(\underline{F}_{CIC,1}^{-1}(q), \overline{y}) - \max(\overline{F}_{CIC,0}^{-1}(q),\underline{y}).
\end{align*}

Finally, we introduce the two following conditions, which ensure that the CIC bounds are well-defined and sharp.

\begin{hyp}\label{hyp:existenceofmoments}
(Existence of moments)

\medskip
$\int |y|d\overline{F}_{CIC,d}(y)<+\infty$ and $\int
|y|d\underline{F}_{CIC,d}(y)<+\infty$ for $d\in \{0,1\}.$
\end{hyp}
\begin{hyp}\label{hyp:increasingbounds}
(Increasing bounds)

\medskip
For $(d,g,t)\in \Supp((D,G,T))$, $F_{Y_{dgt}}$ is continuously
differentiable, with positive derivative on the interior of $\Supp(Y)$. Moreover,
$\underline{T}_d, \overline{T}_d, G_d(\underline{T}_d), G_d(\overline{T}_d), C_d(\underline{T}_d)$ and $C_d(\overline{T}_d)$ are increasing on $\Supp(Y)$.
\end{hyp}

\begin{thm}\label{thm:partial_ident}
Assume that Assumptions \ref{hyp:timeinvariance2}-\ref{hyp:first_stage} are satisfied and $0<P(D_{01}=1)\ne P(D_{00}=1)<1$. Then:
\begin{enumerate}
\item If Assumption \ref{hyp:common_trends_conditional}' holds and $P(\underline{y}\leq Y(d)\leq \overline{y})=1$ for $d\in \{0,1\}$,
$\underline{W}_{TC}\leq \Delta \leq \overline{W}_{TC}.$\footnote{It is not difficult to show that these bounds are sharp. We omit the proof due to a concern for brevity.}
\item If Assumptions \ref{hyp:monotonicity1}-\ref{hyp:existenceofmoments} hold, $F_{Y_{11}(d)|S}(y) \in [
\underline{F}_{CIC,d}(y), \overline{F}_{CIC,d}(y)]$ for $d\in \{0,1\}$, $\Delta \in [\underline{W}_{CIC}, \overline{W}_{CIC}]$ and $\tau_q \in [\underline{\tau}_q,\overline{\tau}_q]$. These bounds are sharp if Assumption \ref{hyp:increasingbounds} holds.
\end{enumerate}
\end{thm}
The reasoning underlying the TC bounds goes as follows. Assume for instance that the treatment rate increases in the control group. Then, the difference between
$E(Y_{101})$ and $E(Y_{100})$ arises both from the trend on $Y(1)$, and from the fact the former expectation is for units treated at $T=0$ and switchers, while the latter is only for units treated at $T=0$. Therefore, we can no longer identify the trend on $Y(1)$ among units treated at $T=0$. But when the outcome has bounded support, this trend can
be bounded, because we know the percentage of the control group switchers account for. A similar reasoning can be used to bound the trend on $Y(0)$ among units untreated at $T=0$. Eventually, $\Delta$ can also be bounded. The smaller the change of the treatment rate in the control group, the tighter the bounds.

\medskip
The reasoning underlying the CIC bounds goes as follows. When $0<P(D_{00}=1)\ne P(D_{01}=1)<1$, the second matching described in Subsection \ref{sec:IVCIC} collapses, because treated (resp. untreated) observations in the control group are no longer comparable in period 0 and 1 as explained in the previous paragraph. Therefore, we cannot match period 0 and period 1 observations on their rank anymore. However, we know the percentage of the control group switchers account for, so we can match period 0 observations to their best- and worst-case rank counterparts in period 1.

\medskip
If the support of the outcome is unbounded, $\underline{F}_{CIC,0}$ and $\overline{F}_{CIC,0}$ are proper cdfs when $\lambda_{00} > 1$, but they are defective when $\lambda_{00} < 1$. On the contrary, $\underline{F}_{CIC,1}$ and $\overline{F}_{CIC,1}$ are always proper cdfs. As a result, when $\Supp(Y)$ is unbounded and $\lambda_{00} > 1$, the CIC bounds we derive for $\Delta$ and $\tau_q$ are finite under Assumption \ref{hyp:existenceofmoments}. The TC bounds, on the other hand, are always infinite when $\Supp(Y)$ is unbounded.

\subsection{Other extensions}
\label{sec:others}

In the supplementary material, we present additional extensions that we discuss briefly here.

\subsubsection*{Multiple groups and multiple periods}

With multiple groups and periods, we show that one can gather groups into ``supergroups'' for each pair of consecutive dates, depending on whether their treatment increases, is stable, or decreases. Then, a properly weighted sum of the estimands for each pair of dates identifies a weighted average of the LATEs of units switching at any point in time.

\subsubsection*{Particular fuzzy designs}

Up to now, we have considered general fuzzy situations where the $P(D_{gt}=d)$ are restricted only by Assumption
\ref{hyp:first_stage}. In our supplementary material, we consider two interesting special cases. First, we show that when $P(D_{00}=1)=P(D_{01}=1)=P(D_{10}=1)=0$,
identification of the average treatment effect on the treated can be obtained under the same assumptions as those of the standard DID or CIC model.
Second, we consider the case where $P(D_{00}=0)=P(D_{01}=0) \in \{0,1\}$. Such situations arise when a policy is extended to a previously ineligible group, or when a program or a technology previously available in some geographic areas is extended to others (see e.g. \citep{field2007}). One can show that Theorem \ref{thm:IV-DID1} still holds in this special case. On the other hand, Theorems \ref{thm:IV-DID2}-\ref{thm:ident_CIC} do not hold. In such instances, identification must rely on the assumption that $Y(0)$ and $Y(1)$ change similarly over time.

\subsubsection*{Including covariates}

We also propose Wald-DID, Wald-TC, and Wald-CIC estimands with covariates. Including covariates in the analysis has two advantages. First, our estimands with covariates rely on conditional versions of our assumptions, which might be more plausible than their unconditional counterparts. Second, there might be instances where $P(D_{00}=d)\ne P(D_{01}=d)$ but
$P(D_{00}=d|X)=P(D_{01}=d|X)>0$ almost surely for some covariates $X$, meaning that in the control group, the change in the treatment rate is driven by a change in the distribution of $X$. If that is the case, one
can use our results with covariates to point identify treatment effects among switchers, while one would only
obtain bounds without covariates.

\subsubsection*{Panel data}

Finally, we discuss the plausibility of our assumptions when panel data are available. Firstly,
Assumption \ref{hyp:timeinvariance2} is well suited for repeated cross sections or cohort data, but less so for panel data.
Then, it implies that within each group units can switch treatment in only one direction, because $V$ does not depend on time.
Actually, this assumption is not necessary for our results to hold. For instance, Theorem \ref{thm:IV-DID1} still holds if Assumptions \ref{hyp:timeinvariance2} and \ref{hyp:common_trends_within} are replaced by
\begin{equation}
D_{it} =1\{V_{it} \geq v_{G_it}\}, \quad \text{with } V_{i1}| G_i \sim V_{i0}|G_i
\label{eq:select_panel}
\end{equation}
and
\begin{equation*}
E(Y_{i1}(1)-Y_{i1}(0)|G_i, V_{i1}\geq v_{G_i0} )=E(Y_{i0}(1)-Y_{i0}(0)|G_i, V_{i0}\geq v_{G_i0}),
\end{equation*}
where we index random variables by $i$ to distinguish individual effects from constant terms. The result applies to treatment and control group switchers, respectively defined as $S_i=\{V_{i1} \in [v_{11}, v_{10}), G_i=1\}$ and $S'_i=\{V_{i1} \in [\min(v_{01}, v_{00}),\max(v_{01}, v_{00})), G_i=0\}$. Theorems \ref{thm:IV-DID2} and \ref{thm:ident_CIC} also still hold if Assumption \ref{hyp:timeinvariance2} is replaced by Equation \eqref{eq:select_panel}, under modifications of Assumptions \ref{hyp:common_trends_conditional}' and \ref{hyp:monotonicity1} that we present in our supplementary material.

\medskip
Secondly, we discuss the validity of our estimands under Equation \eqref{eq:select_panel} and the following model:
\begin{equation*}
Y_{it}=\Lambda\left(\alpha_i + \gamma_t + \left[\beta_i + \lambda_t\right] D_{it}+\eps_{it}\right),
\end{equation*}
with
\begin{eqnarray*}
&&(\alpha_i,\beta_i)|G_i,V_{i1}\geq v_{G_i0}\sim (\alpha_i,\beta_i)|G_i,V_{i0}\geq v_{G_i0},\\
&&(\alpha_i,\beta_i)|G_i,V_{i1}< v_{G_i0}\sim (\alpha_i,\beta_i)|G_i,V_{i0}< v_{G_i0},
\end{eqnarray*}
and $\Lambda(.)$ strictly increasing. We prove that $\Delta$ is identified by the Wald-DID, Wald-TC, or Wald-CIC estimand under alternative restrictions on $\Lambda(.)$,
$\lambda_t$, and the distribution of $\eps_{it}$.

\section{Estimation and inference}\label{sec:inference}

In this section, we study the asymptotic properties of the estimators corresponding to the estimands introduced in Section \ref{sec:ident}, assuming we have an i.i.d. sample with the same distribution as $(Y,D,G,T)$.
\begin{hyp}\label{hyp:iid}
(Independent and identically distributed observations)

\medskip
$(Y_i,D_i,G_i,T_i)_{i=1,...,n}$ are i.i.d.
\end{hyp}
Let $\mathcal{I}_{gt}=\{i: G_i=g, T_i=t\}$ (resp.  $\mathcal{I}_{dgt}=\{i: D_i=d,G_i=g, T_i=t\}$) and $n_{gt}$ (resp. $n_{dgt}$) denote the size
of $\mathcal{I}_{gt}$ (resp. $\mathcal{I}_{dgt}$) for all $(d,g,t)\in\{0,1\}^3$. For $d\in \{0,1\}$, let
$$\widehat{\delta}_d = \frac{1}{n_{d01}} \sum_{i\in \mathcal{I}_{d01}} Y_i - \frac{1}{n_{d00}} \sum_{i\in \mathcal{I}_{d00}} Y_i.$$ Let
\begin{align*}
\widehat{W}_{DID} &=  \frac{\frac{1}{n_{11}} \sum_{i\in \mathcal{I}_{11}} Y_i - \frac{1}{n_{10}} \sum_{i\in \mathcal{I}_{10}} Y_i
- \frac{1}{n_{01}} \sum_{i\in \mathcal{I}_{01}} Y_i + \frac{1}{n_{00}} \sum_{i\in \mathcal{I}_{00}} Y_i}{\frac{1}{n_{11}} \sum_{i\in \mathcal{I}_{11}} D_i - \frac{1}{n_{10}} \sum_{i\in \mathcal{I}_{10}} D_i
- \frac{1}{n_{01}} \sum_{i\in \mathcal{I}_{01}} D_i + \frac{1}{n_{00}} \sum_{i\in \mathcal{I}_{00}} D_i}, \\
\widehat{W}_{TC} &=  \frac{\frac{1}{n_{11}} \sum_{i\in \mathcal{I}_{11}} Y_i - \frac{1}{n_{10}} \sum_{i\in \mathcal{I}_{10}} \left[Y_i + \widehat{\delta}_{D_i}\right]}{\frac{1}{n_{11}} \sum_{i\in \mathcal{I}_{11}} D_i - \frac{1}{n_{10}} \sum_{i\in \mathcal{I}_{10}} D_i}
\end{align*}
denote our Wald-DID and Wald-TC estimators.

\medskip
Let $\widehat{F}_{Y_{dgt}}$ denote the empirical cdf of $Y$ on the subsample
$\mathcal{I}_{dgt}$, $\widehat{F}_{Y_{dgt}}(y) = \sum_{i\in\mathcal{I}_{dgt}}\mathds{1}\{Y_i\leq y\}/n_{dgt}$. Let $\widehat{F}^{-1}_{Y_{dgt}}(q) =
\inf \{y: \widehat{F}_{Y_{dgt}}(y)\geq q\}$ denote the empirical quantile of order $q\in (0,1)$ of $Y_{dgt}$. Let $\widehat{Q}_d =\widehat{F}_{Y_{d01}}^{-1} \circ \widehat{F}_{Y_{d00}}$ denote the estimator of the quantile-quantile transform. Let
$$\widehat{W}_{CIC}= \frac{\frac{1}{n_{11}}\sum_{i\in \mathcal{I}_{11}} Y_i
- \frac{1}{n_{10}}\sum_{i\in \mathcal{I}_{10}} \widehat{Q}_{D_i}(Y_i)}{\frac{1}{n_{11}}\sum_{i\in
\mathcal{I}_{11}}
D_i - \frac{1}{n_{10}}\sum_{i\in \mathcal{I}_{10}} D_i}$$ denote the Wald-CIC estimator.
Let $\widehat{P}(D_{gt}=d)$ denote the proportion of units with $D=d$ in the sample $\mathcal{I}_{gt}$, let
$\widehat{H}_d=\widehat{F}_{Y_{d10}} \circ \widehat{F}_{Y_{d00}}^{-1}$, and let
$$\widehat{F}_{Y_{11}(d)|S} = \frac{\widehat{P}(D_{10}=d)\widehat{H}_d \circ \widehat{F}_{Y_{d01}}-\widehat{P}(D_{11}=d) \widehat{F}_{Y_{d11}}}{\widehat{P}(D_{10}=d)-\widehat{P}(D_{11}=d)}.$$
Finally, let
$$\widehat{\tau}_q = \widehat{F}_{Y_{11}(1)|S}^{-1}(q) - \widehat{F}_{Y_{11}(0)|S}^{-1}(q)$$
denote the estimator of the LQTE of order $q$ for switchers.

\medskip
We derive the asymptotic behavior of our CIC estimators under the following assumption, which is similar to the one made by \cite{Athey06} for the CIC estimators in sharp settings.

\begin{hyp}
    \label{hyp:techconditioninference1}
    (Regularity conditions for the CIC estimators)

\medskip
$\Supp(Y)$ is a bounded interval $[\underline{y},\overline{y}]$. Moreover, for all $(d,g,t)\in
\{0,1\}^3$, $F_{Y_{dgt}}$ and $F_{Y_{11}(d)|S}$  are continuously differentiable with strictly
positive derivatives on $[\underline{y},\overline{y}]$.
\end{hyp}

Theorem \ref{thm:as_nor_point} below shows that all our estimators are root-n consistent and asymptotically normal. We also derive the influence functions of our estimators. However, because these influence functions take complicated expressions, using the bootstrap might be convenient for inference. For any statistic $T$, we let $T^b$ denote its bootstrap counterpart. For any root-n consistent statistic $\widehat{\theta}$ estimating consistently $\theta$, we say that the bootstrap is consistent if with probability
one and conditional on the sample, $\sqrt{n}(\widehat{\theta}^b - \widehat{\theta})$ converges to the same distribution as the limit distribution of $\sqrt{n}(\widehat{\theta} - \theta)$ (see, e.g., \citep{vanderVaart00}, Section 23.2.1, for a formal definition of conditional convergence). Theorem \ref{thm:as_nor_point} implies that bootstrap confidence intervals are asymptotically valid for all our estimators.

\begin{thm}\label{thm:as_nor_point}
Suppose that Assumptions \ref{hyp:timeinvariance2}-\ref{hyp:first_stage}, \ref{hyp:stable_control}, and
\ref{hyp:iid} hold. Then
\begin{enumerate}
\item If $E(Y^2)<\infty$ and
Assumptions \ref{hyp:common_trends_between}-\ref{hyp:common_trends_within} also hold,
$$\sqrt{n}\left(\widehat{W}_{DID} - \Delta\right) \convL \mathcal{N}\left(0,V\left(\psi_{DID}\right)\right),$$
where $\psi_{DID}$ is defined in Equation \eqref{eq:psi_DID} in the appendix. Moreover, the bootstrap is consistent for $\widehat{W}_{DID}$.
\item  If
$E(Y^2)<\infty$ and
Assumption \ref{hyp:common_trends_conditional}' also holds,
$$\sqrt{n}\left(\widehat{W}_{TC} - \Delta\right) \convL \mathcal{N}\left(0,V\left(\psi_{TC}\right)\right)$$
where $\psi_{TC}$ is defined in Equation \eqref{eq:psi_TC} in the appendix. Moreover, the bootstrap is consistent for $\widehat{W}_{TC}$.
\item If Assumptions \ref{hyp:monotonicity1}, \ref{hyp:data1} and \ref{hyp:techconditioninference1} also hold,
\begin{align*}
\sqrt{n}\left(\widehat{W}_{CIC} - \Delta\right) & \convL \mathcal{N}\left(0,V\left(\psi_{CIC}\right)\right), \\
\sqrt{n}\left(\widehat{\tau_q} - \tau_q\right) & \convL \mathcal{N}\left(0,V\left(\psi_{q,CIC}\right)\right),
\end{align*}
where $\psi_{CIC}$ and $\psi_{q, CIC}$ are defined in Equations \eqref{eq:psi_W_CIC} and \eqref{eq:psi_tau_CIC} in the appendix. Moreover, the bootstrap is consistent for both estimators.
\end{enumerate}
\end{thm}
The result is straightforward for $\widehat{W}_{DID}$ and $\widehat{W}_{TC}$. Regarding $\widehat{W}_{CIC}$ and $\widehat{\tau_q}$, our proof differs from the one of \cite{Athey06}. It is based on the weak convergence of the empirical cdfs of the different subgroups, and on a repeated use of the functional delta method. This approach can be readily applied to other functionals of $(F_{Y_{11}(0)|S},F_{Y_{11}(1)|S})$.

\medskip
In our supplementary material, we extend the asymptotic theory presented here in several directions. Firstly, we show that we can allow for clustering. Even with repeated cross section or cohort data, independence is a strong assumption in DID analysis: clustering at the group level can induce both cross-sectional and serial correlation within clusters (see e.g. \citep{Bertrand04}).  Secondly, we consider estimators of the bounds presented in Subsection \ref{sec:partial} and we derive their limiting distributions. Thirdly, we consider estimators incorporating covariates and we also derive their limiting distributions.


\section{Application: returns to schooling in Indonesia}
\label{sec:applications2}

\subsection{Results with the same control  and treatment groups as in \cite{Duflo01}}\label{sec:application_oldresults}

In 1973-1974, the Indonesian government launched a major primary school construction program, the so-called INPRES program. \cite{Duflo01} uses it to measure returns to education among men. Children born between 1968 and 1972 (cohort 1) entered primary school after the program was launched, while children born between 1957 and 1962 (cohort 0) had finished primary school by that time. The author constructs two ``supergroups'' of districts, by regressing the number of primary schools constructed on the number of school-age children in each district. She defines treatment districts as those with a positive residual in that regression. She starts by using a simple Wald-DID with her two groups of districts and cohorts to estimate returns to education. She also estimates a 2SLS regression of wages on cohort dummies, district dummies, and years of schooling, using the interaction of cohort 1 and schools constructed in one's district of birth as the instrument for years of schooling.

\medskip
As an alternative, we apply our results to the author's data. Because years of schooling changed between cohorts 0 and 1 in her control group, we estimate bounds for returns to schooling. These bounds are similar to those presented in Subsection \ref{sec:partial}, but account for the fact that schooling is not binary (see Subsection \ref{sub:bounds_appli} of the supplementary material for more details). We estimate TC bounds relying on Assumptions \ref{hyp:timeinvariance2}-\ref{hyp:first_stage} and \ref{hyp:common_trends_conditional}', and CIC bounds relying on Assumptions \ref{hyp:timeinvariance2}-\ref{hyp:first_stage} and \ref{hyp:monotonicity1}-\ref{hyp:data1}. Because the support of wages does not have natural boundaries, we use the lowest and highest wage in the sample.

\medskip
Results are shown in Table \ref{table_returns_Duf_groups}.\footnote{Our Wald-DID and 2SLS estimates differ slightly from those of in \cite{Duflo01} because we were not able to obtain exactly her sample of 31,061 observations.} The Wald-DID is large. However, it is not significantly different from $0$, which is the reason why the author turns to a 2SLS regression with cohort and district dummies. The estimate of returns to schooling in that regression is equal to 7.3\% and is more precisely estimated.\footnote{This point estimate was significant at the 5\% level in the original paper (see the 3rd line and 1st column of Table 7 in \citep{Duflo01}). But once clustering standard errors at the district level, which has become standard practice in DID analysis since \cite{Bertrand04}, it loses some statistical significance.}

\begin{table}[H]
\begin{center}
\caption{Returns to education using the groups in \cite{Duflo01}}
\begin{tabular}{lcc}
\hline \hline \\
 &  Estimate & 95\% CI \\
Wald-DID & 0.195  &  [-0.102, 0.491] \\
2SLS with fixed effects & 0.073 & [-0.011, 0.157]\\
TC bounds  &  [-3.70, 2.18]   &[-5.29, 3.00]  \\
CIC bounds & [-5.60, 3.36] & [-8.00, 4.63] \\
\\ \hline \hline
\end{tabular}\label{table_returns_Duf_groups}
\end{center}
\footnotesize{\textit{Notes}. Sample size: 30 828 observations. Confidence intervals account for clustering at the district level.}
\end{table}

The Wald-DID and 2SLS estimates rely on the assumption that returns to education are homogeneous between districts. The INPRES program explains a small fraction of the differences in increases in schooling between districts. A district-level regression of the increase in years of schooling between cohort 0 and 1 on primary schools constructed
has an R-squared of 0.03 only.  Accordingly, years of schooling increased almost as much in the author's control group than in her treatment group: while the average of years of schooling increased by 0.47 between cohort 0 and 1 in her treatment group, it increased by 0.36 in her control group (see Table 3 in \citep{Duflo01}). Therefore, one can show that under Assumptions \ref{hyp:timeinvariance2}-\ref{hyp:common_trends_between} and \ref{hyp:common_trends_within}, the author's Wald-DID is equal to $0.47/0.11 \times ACR-0.36/0.11 \times ACR',$ where $ACR$ and $ACR'$ respectively denote the ACR parameters we introduced in Section \ref{sec:multivariate_treatment} in her treatment and control groups. If $ACR \ne ACR'$, this Wald-DID could lie far from both $ACR$ and $ACR'$. Moreover, \cite{deChaisemartin16} show that under Assumptions \ref{hyp:timeinvariance2}-\ref{hyp:common_trends_between} and \ref{hyp:common_trends_within}, the author's 2SLS regression with fixed effects estimates a weighted sum of switchers' returns to education across districts, with potentially many negative weights. We estimate the weights, and find that almost half of districts receive a negative weight in this regression, with negative weights summing up to -3.28. Here again, if switchers' returns are heterogeneous across districts, this 2SLS coefficient could lie far from returns in any district.

\medskip
However, returns to schooling might differ across districts. In cohort 0, years of schooling were substantially higher in control than in treatment districts (see Table 3 in \citep{Duflo01}). This difference in years of schooling might for instance indicate a higher level of economic development in control districts, in which case demand for skilled labor and returns to education could be higher there.

\medskip
Our TC and CIC bounds do not rely on this assumption. But because years of schooling changed substantially in the authors' control group, they turn out not to be informative. One could argue that this is due to outliers or measurement error, because we use the minimum and maximum wages in the sample as estimates of the boundaries of the support of wages. Using instead the first and third quartile of wages still yields very uninformative bounds: our TC and CIC bounds are respectively equal to [ -0.79, 0.36] and [ -0.97, 0.35].

\medskip
Overall, using the treatment and control groups of districts defined in \cite{Duflo01} either yields point estimates relying on a questionable assumption, or wide and uninformative bounds.

\subsection{Results with new control and treatment groups}\label{sec:application_newresults}

In this subsection, we follow the results of Subsection \ref{sec:multiple_time_periods} and form three supergroups of districts depending of whether their years of schooling increased, remained stable, or decreased between cohorts 0 and 1. This approach enables us to obtain point estimates of returns to schooling, without assuming that returns are homogeneous between districts or over time. A difficulty, however, is that the author's data set is a survey including only a subset of the population of each district. Therefore, we cannot identify directly these supergroups, but we need to estimate them. Studying how this first-step estimation should be accounted for in our second-step estimation of returns to schooling goes beyond the scope of this paper. Therefore, what follows is tentative. In Subsection \ref{sub:robustness} of our supplementary material, we still present various robustness checks suggesting that our second-step estimates should not change much if we were to account for this first-step estimation.\footnote{There are many other applications of the fuzzy DID method where the set of groups where treatment is stable is known and does not need to be estimated. Examples include \cite{field2007} or \cite{gentzkow2011}, which we revisit in our supplementary material.}

\medskip
The procedure we use to estimate the supergroups should classify as controls only districts with a stable distribution of education. Any classification method leads us to make two types of errors: classify some districts where the distribution of education remained constant as treatments (type 1 error); and classify some districts where this distribution changed as controls (type 2 error). Type 1 errors are innocuous. For instance, if Assumptions \ref{hyp:common_trends_between} and \ref{hyp:common_trends_within} are satisfied, all control districts have the same  evolution of their expected outcome. Misclassifying some as treatment districts leaves our estimators unchanged, up to sampling error. On the other hand, type 2 errors are a more serious concern. They lead us to include districts where the true distribution of education was not stable in our super control group, thus violating one of the requirements of Theorem \ref{thm:manygroups}. We therefore choose a method based on chi-squared tests with very liberal level. Specifically, we assign a district to our control group if the p-value of a chi-squared test comparing the distribution of education between the two cohorts in that district is greater than 0.5.

\medskip
We end up with control ($G^*=0$) and treatment groups ($G^*=1$ and $G^*=-1$) respectively made up of 64, 123, and 97 districts.  Table \ref{table_first_stage} shows that in treatment districts where schooling increased, cohort 1 completed one more year of schooling than cohort 0. In treatment districts where schooling decreased, cohort 1 completed nine months less of schooling than cohort 0. Finally, in control districts, the number of years of schooling did not change.

\begin{table}[H]
\begin{center}
\caption{Years of schooling completed in the new groups of districts}
\begin{tabular}{l c c c c c}
\hline \hline
\\
 & & Cohort 0 & Cohort 1 & Evolution & (s.e.) \\
Districts where schooling increased & ($G^*=1$) & 8.65 & 9.64 & 0.99 & (0.082) \\
Control districts & ($G^*=0$) & 9.60  & 9.55 & -0.05 & (0.097) \\
Districts where schooling decreased & ($G^*=-1$) & 10.17 & 9.43 & -0.74 & (0.080) \\
\\ \hline \hline
\end{tabular}\label{table_first_stage}
\end{center}
\footnotesize{\textit{Notes}. Sample size: 30 828 observations. Standard errors are clustered at the district level.}
\end{table}

We now follow results from Theorem \ref{thm:manygroups} to estimate returns to education. In Table \ref{table_returns}, we report the Wald-DID, Wald-TC, and Wald-CIC estimates with our new groups of districts.\footnote{Our Wald-DID estimate is actually a weighted average of the Wald-DID estimate comparing $G^*=1$ and $G^*=0$ and of the Wald-DID estimate comparing $G^*=-1$ and $G^*=0$. The same holds for the Wald-TC and Wald-CIC estimates. The formulas of the corresponding estimands are given in Theorem \ref{thm:manygroups}. Also, to estimate the numerators of the Wald-CICs, we group schooling into 5 categories (did not complete primary school, completed primary school, completed middle school, completed high school, completed college). Thus, we avoid estimating the $Q_d$s on small numbers of units. To be consistent, we also use this definition to estimate the numerators of the Wald-TCs. Using years of schooling hardly changes our Wald-TC estimate.} The Wald-DID is large, and suggests returns of 14\% per year of schooling. Our Wald-TC and Wald-CIC estimators are substantially smaller, around 10\% per year of schooling. They significantly differ from the Wald-DID, with t-stats respectively equal to -4.27 and -4.61.

\begin{table}[H]
\begin{center}
\caption{Returns to education using our new groups}
\begin{tabular}{l c c c c}
\hline \hline \\
   & $W_{DID}$ & $W_{TC}$ & $W_{CIC}$ \\
Returns to education  & 0.140 & 0.101 & 0.099  \\
   & (0.015) & (0.017) & (0.017) \\
\\ \hline \hline
\end{tabular}\label{table_returns}
\end{center}
\footnotesize{\textit{Notes}. Sample size: 30 828 observations. Standard errors are clustered at the district level.}
\end{table}
We now follow Subsection \ref{sec:comparison} to choose between our estimators.
We start by conducting placebo tests, which are presented in Subsection \ref{sub:placebos} in the supplementary material. In our three groups of districts, men born between 1945
and 1950 (cohort -2) did not complete a significantly different number of years of schooling from men born between 1951 and 1956 (cohort -1). Likewise, in $G^*=0$ and $G^*=1$ the number of years of schooling did not significantly change between cohorts -1 and 0. This enables us to test whether the assumptions underlying our estimators are satisfied for these older cohorts, as explained in Subsection \ref{sec:comparison}. The placebo DID comparing the evolution of wages from cohort -2 to -1 in $G^* = 0$ and $G^* = 1$ is positive and close to being significant (t-stat=1.43). All the other placebo DID estimators (between cohort -1 and 0, or between $G^* = 0$ and $G^* =-1$) are small and insignificant. All the placebo Wald-TC and Wald-CIC estimators are small and insignificant. Overall, our placebo analysis lends slightly stronger support to our Wald-TC and Wald-CIC estimator, even though the fact that the placebo DID between cohort -2 and -1 is not truly significant prevents us from making definitive conclusions.

\medskip
Instead, our choice must be based on economic theory and a careful discussion of the assumptions underlying each estimator. The Wald-DID relies on Assumption \ref{hyp:common_trends_within}, which might not be plausible in this context. Rearranging the equation in Assumption \ref{hyp:common_trends_within},  one can see that it implies, for instance, that the wage gap between high-school graduates in cohort 0 and 1 should have remained the same if they had only completed primary school.  Had they only completed primary school, high-school graduates would have joined the labor market earlier, and would have had more labor market experience at the time we compare their wages. The wage gap between the two cohorts might then have been lower, because returns to experience tend to decrease (see e.g. \citep{mincer1979}). The data supports this hypothesis. In the control group, while high-school graduates in cohort 1 earn 54\% less than their cohort 0 counterpart, the gap is only 20\% for non-graduates, and the difference
is significant (t-stat=-7.64). This difference could arise from differences between the two populations: the cohort gap among non-graduates might not be equal to the cohort gap we would have observed among graduates had they not graduated. Still, it seems unlikely that differences between the two populations can fully account for this almost threefold difference. We also explain in Subsection \ref{sub:placebos} of the  supplementary material why decreasing returns to experience can account for the relatively small placebo DIDs, as they would lead to smaller violations of Assumption \ref{hyp:common_trends_within} for the older cohorts we compare in those placebos. Overall, Assumption \ref{hyp:common_trends_within} is not warranted in this context. We therefore choose the Wald-TC and Wald-CIC as our preferred estimators, because they do not rely on this assumption.



\medskip
Finally, in Subsection \ref{sub:robustness} in the supplementary material we show robustness checks suggesting that our estimates of returns to schooling should not change much if we were to account for the first-step estimation of the supergroups. First, we show that our results are robust to our choice of the p-value defining the treatment and the control group. Defining control districts as those where the p-value of the chi-squared test comparing the distribution of schooling between the two cohorts is greater than 0.6 significantly changes the composition of our three groups but does not affect our results. Second, we show that the fact we use the data twice - to construct our groups and to measure returns to schooling - does not seem to bias our estimators. Using a split-sample procedure whereby half of the sample is used to construct our groups and the other half is used to estimate returns to schooling yields similar results. If anything, this procedure produces slightly larger point estimates than those we report here. Third, we give suggestive evidence that while accounting for the first-step estimation would increase the standard errors of our estimates of returns to schooling, our main conclusions would remain unaffected.

\section{Conclusion}

In many applications of the DID method, the treatment increases more in the treatment group, but some units are also treated in the control group, and some units remain untreated in the treatment group. In such fuzzy designs, a popular estimand of treatment effects is the DID of the outcome divided by the  DID of the treatment. We first show that this  ratio identifies the LATE of treatment group switchers only if two restrictions on treatment effect heterogeneity are satisfied, in addition to the usual common trend assumption. The LATE of units treated at both dates must not change over time. Moreover, when the share of treated units varies in the control group, the LATEs of treatment and control group switchers must be equal. Second, we propose two new estimands that can be used when the share of treated units in the control group is stable, and that do not rely on any treatment effect homogeneity assumption. We use our results to revisit \cite{Duflo01}.

\medskip
When a policy is extended to a previously ineligible subgroup or when the treatment is assigned at the group level, control groups whose exposure to the treatment does not change over time are usually readily available. Examples include \cite{field2007} or \cite{gentzkow2011}, which we revisit in our supplementary material. When the treatment is assigned at the individual level and no policy rule warrants that treatment remains stable in some groups, such control groups usually still exist but they need to be estimated. This is
the case in \cite{Duflo01}. In future work, we will study how researchers should account for this first-step estimation when estimating treatment effects. Finally, when researchers cannot find a control group whose exposure to the treatment is stable over time, our results show that their conclusions will rest on treatment effect homogeneity assumptions. These assumptions should then be discussed.

\newpage
\bibliography{biblio}
\normalsize

\newpage
\appendix

\section*{Main proofs}

The lemmas prefixed by S are stated and proven in our supplementary material (see \citep{deChaisemartin15}). To simplify the notation, throughout the proofs we adopt the following normalization, which is without loss of generality: $v_{10}=v_{00}$. For any $(d,g,t)\in \Supp((D,G,T))$, we also let $p_{gt}=P(G=g,T=t)$, $p_{dgt}=P(D=d,G=g,T=t)$, $p_{d|gt}=P(D_{gt}=d)$, and $F_{dgt}=F_{Y_{dgt}}$. Finally, for any $\Theta \subset \mathbb{R}^k$, let $\interior \Theta$ denote its interior and let $\mathcal{C}^0(\Theta)$ and $\mathcal{C}^1(\Theta)$ denote respectively the set of continuous functions and the set of continuously
differentiable functions with strictly positive derivative  on $\Theta$. We often use this notation with $\Theta=\Supp(Y)$, in which case we respectively denote these sets by $\mathcal{C}^0$ and $\mathcal{C}^1$.

\subsection*{Theorem \ref{thm:IV-DID1}}

\medskip
\textbf{Proof of 1 when $p_{1|01} \geq p_{1|00}$.}

\medskip
It follows from Assumptions \ref{hyp:timeinvariance2} and \ref{hyp:first_stage} that
\begin{eqnarray}\label{eq:change_treatment_rate}
p_{1|11}-p_{1|10}&=&P(V\geq v_{11}|G=1,T=1)-P(V\geq v_{00}|G=1,T=0)\nonumber\\
&=&P(S|G=1).
\end{eqnarray}
Moreover,
\begin{eqnarray}
&&E(Y_{11})-E(Y_{10})\nonumber\\
&=&E\left(Y_{11}(1)\left[\mathds{1}\{V_{11}\in [v_{11}, v_{00})\} + \mathds{1}\{V_{11}\geq v_{00}\}
\right]\right) + E(Y_{11}(0)\mathds{1}\{V_{11}< v_{11}\}) \nonumber\\
& - &   E(Y_{10}(1)\mathds{1}\{V_{10}\geq v_{00}\}) - E(Y_{10}(0)\mathds{1}\{V_{10}< v_{00}\}) \nonumber\\
& =& E(Y_{11}(1)-Y_{11}(0)|S) P(S|G=1) + E(Y_{11}(0))- E(Y_{10}(0)) \nonumber\\
& + & E\left((Y_{11}(1)-Y_{11}(0))\mathds{1}\{V_{11}\geq v_{00}\}\right) -  E\left((Y_{10}(1)-Y_{10}(0))\mathds{1}\{V_{10}\geq v_{00}\}\right) \nonumber\\
& =& \Delta P(S|G=1) + E(Y_{11}(0))- E(Y_{10}(0))  \label{eq:proof_WDID1}
\end{eqnarray}
The first equality follows from Assumptions \ref{hyp:timeinvariance2} and \ref{hyp:first_stage}, the second follows from Assumption \ref{hyp:timeinvariance2}, and the third follows from Assumptions \ref{hyp:timeinvariance2} and \ref{hyp:common_trends_within}.

Similarly, one can show that
\begin{eqnarray}
p_{1|01}-p_{1|00}&=&P(S'|G=0) \label{eq:change_treatment_rate_control}\\
E(Y_{01})-E(Y_{00}) &=& \Delta' P(S'|G=0) + E(Y_{01}(0))- E(Y_{00}(0)). \label{eq:proof_WDID_control}
\end{eqnarray}
Taking the difference between Equations \eqref{eq:proof_WDID1} and \eqref{eq:proof_WDID_control}, and using Assumption \ref{hyp:common_trends_between}, we obtain
$$DID_Y = \Delta P(S|G=1) - \Delta' P(S'|G=0).$$
Dividing each side by $DID_D$ and using Equations \eqref{eq:change_treatment_rate} and \eqref{eq:change_treatment_rate_control} yields the result.

\medskip
\textbf{Proof of 1 when $p_{1|01} < p_{1|00}$}

\medskip
In this case, reasoning similarly as in the derivation of Equation \eqref{eq:change_treatment_rate} yields
\begin{eqnarray}
p_{1|00}-p_{1|01}&=&P(S'|G=0). \label{eq:change_treatment_rate2}
\end{eqnarray}
Moreover,
\begin{eqnarray}
&&E(Y_{01})-E(Y_{00})\nonumber\\
&=&E\left(Y_{01}(1)\left[\mathds{1}\{V_{01} \geq v_{00}\} - \mathds{1}\{V_{01} \in [v_{00}, v_{01})\}\right]
\right) + E\left(Y_{01}(0)\times \right. \nonumber \\
&& \left.\left[\mathds{1}\{V_{01} \in [v_{00}, v_{01})\}+ \mathds{1}\{V_{01} < v_{00}\}\right]\right)
- E\left(Y_{00}(1)\mathds{1}\{V_{00} \geq v_{00}\}\right)
- E\left(Y_{00}(0)\mathds{1}\{V_{00} < v_{00}\}\right) \nonumber \\
& = & - \Delta' P(S'|G=1) + E(Y_{01}(0) - Y_{00}(0)) \nonumber \\
& + & E\left((Y_{01}(1)-Y_{01}(0)\mathds{1}\{V_{01} \geq v_{00}\}\right) -
E\left((Y_{00}(1)-Y_{00}(0)\mathds{1}\{V_{00} \geq v_{00}\}\right) \nonumber \\
& = & - \Delta' P(S'|G=1) + E(Y_{01}(0) - Y_{00}(0)).
\label{eq:proof_WDID2}
\end{eqnarray}
The first equality follows from Assumption \ref{hyp:timeinvariance2} and $p_{1|01} < p_{1|00}$, the second follows from
Assumption \ref{hyp:timeinvariance2}, and the third follows from Assumptions \ref{hyp:timeinvariance2} and \ref{hyp:common_trends_within}.
Taking the difference between Equations \eqref{eq:proof_WDID1} and \eqref{eq:proof_WDID2}, and using Assumption \ref{hyp:common_trends_between}, we obtain
$$DID_Y  = \Delta P(S|G=1)  + \Delta' P(S'|G=0).$$
Dividing each side of the previous display by $DID_D$ and using Equations \eqref{eq:change_treatment_rate} and \eqref{eq:change_treatment_rate2} yields the result.

\medskip
\textbf{Proof of 2}

The result follows directly from the first point of the theorem.

\medskip
\textbf{Bias term without Assumption \ref{hyp:common_trends_within}}

Using the Equations above \eqref{eq:proof_WDID1} and \eqref{eq:proof_WDID2}, we obtain that when Assumption \ref{hyp:stable_control} holds but
Assumption \ref{hyp:common_trends_within} does not,
$$W_{DID} = \Delta + \frac{1}{DID_D}\left((\Delta_{11} - \Delta_{10}) p_{1|10} - (\Delta_{01} - \Delta_{00})p_{1|00}
\right),$$
where $\Delta_{gt}=E\left(Y_{gt}(1)-Y_{gt}(0)|D(0)=1\right) \;_\Box$

\subsection*{Theorem \ref{thm:IV-DID2}}

Following the same steps as those used to derive Equation \eqref{eq:proof_WDID1}, we obtain
\begin{eqnarray}\label{eq:proof_WTC1}
& & E(Y_{11})-E(Y_{10}) \nonumber \\
& = & E(Y_{11}(1)-Y_{11}(0)|S)P(S|G=1) + E\left(Y_{11}(1)-Y_{10}(1)|G=1,V\geq v_{00}\right)P(V\geq v_{00}|G=1)\nonumber\\
&+&E\left(Y_{11}(0)-Y_{10}(0)|G=1,V< v_{00}\right)P(V< v_{00}|G=1).
\end{eqnarray}
Then,
\begin{eqnarray}\label{eq:proof_WTC2}
\delta_1
&=&E(Y_{01}(1)|G=0,V\geq v_{01})-E(Y_{00}(1)|G=0,V\geq v_{00})\nonumber\\
&=&E(Y_{01}(1)-Y_{00}(1)|G=0,V\geq v_{00}).
\end{eqnarray}The first equality follows from Assumption \ref{hyp:timeinvariance2}. The second one follows from the fact that $p_{1|01}=p_{1|00}$ combined with Assumption \ref{hyp:timeinvariance2} implies that $\{G=0,V\leq v_{01}\}=\{G=0,V\leq v_{00}\}$.

\medskip
Similarly,
\begin{equation}\label{eq:proof_WTC3}
\delta_0=E(Y_{01}(0)-Y_{00}(0)|G=0,V< v_{00}).
\end{equation}
Finally, the result follows combining Equations \eqref{eq:proof_WTC1}, \eqref{eq:proof_WTC2}, \eqref{eq:proof_WTC3},  and Assumption \ref{hyp:common_trends_conditional}', once noted that $p_{1|10}=P(V\geq v_{00}|G=1)$ and $P(S|G=1)=p_{1|11}-p_{1|10} \; {}_\Box$

\subsection*{Theorem \ref{thm:ident_CIC}}

We first prove the following result, which holds irrespective of whether Assumption \ref{hyp:stable_control} holds:
\begin{equation}
F_{Y_{11}(d)|S} =\frac{p_{d|11}F_{d11}-p_{d|10}H_d \circ \left(\lambda_{0d}F_{d01}+(1-\lambda_{0d})F_{Y_{01}(d)|S'}\right)}{p_{d|11}-p_{d|10}},
\label{eq:F_CIC}
\end{equation}
where  $\lambda_{gd}=p_{d|g1}/p_{d|g0}$. We establish \eqref{eq:F_CIC} for $d=0$ only, the reasoning is similar for $d=1$.

First,
\begin{eqnarray*}
P(S|G=1, T=1, V<v_{00}) & = & \frac{P(S|G=1)}{P(V<v_{00}|G=1,T=0)} \\
& = & \frac{p_{0|10}- p_{0|11}}{p_{0|10}}.
\end{eqnarray*}
The first equality stems from Assumption \ref{hyp:timeinvariance2},  the second from
Equation \eqref{eq:change_treatment_rate} and Assumption \ref{hyp:timeinvariance2}. Moreover,
\begin{eqnarray*}
    F_{Y_{11}(0)|V<v_{00}}
	& = & P(S|G=1, T=1, V<v_{00})  F_{Y_{11}(0)|S} \\
	& + &  (1-P(S|G=1, T=1, V<v_{00})) F_{Y_{11}(0)|V<v_{11}} \\
& = & \frac{p_{0|10}- p_{0|11}}{p_{0|10}}F_{Y_{11}(0)|S} + \frac{p_{0|11}}{p_{0|10}}F_{011}.
\end{eqnarray*}
Therefore,
\begin{equation}
F_{Y_{11}(0)|S} = \frac{p_{0|11}F_{011}-p_{0|10}F_{Y_{11}(0)|V<v_{00}}}{p_{0|11}-p_{0|10}}.
\label{eq_1}
\end{equation}
Then, we show that for all $y \in \Supp(Y_{11}(0)|V<v_{00})$,
\begin{equation}
    F_{Y_{11}(0)|V<v_{00}}= F_{010} \circ F^{-1}_{000} \circ F_{Y_{01}(0)|V<v_{00}}.
    \label{eq_2}
\end{equation}
For all $(g,t)\in\{0,1\}^2$,
\begin{eqnarray*}
    F_{Y_{gt}(0)|V<v_{00}}(y) &= & P(h_0(U_{0},t) \leq y |G=g,T=t, V<v_{00}) \\
    & = & P(U_{0} \leq h_0^{-1}(y,t) | G=g,V<v_{00}) \\
    & = & F_{U_{0}|G=g,V<v_{00}}( h_0^{-1}(y,t)),
\end{eqnarray*}
where the first and second equalities follow from Assumption \ref{hyp:monotonicity1}.
Assumptions \ref{hyp:monotonicity1}  and \ref{hyp:data1} imply that
$F_{U_{0}|G=g,V<v_{00}}$ is strictly increasing. Hence, its inverse exists and for all $q \in
(0,1)$,
$$F_{Y_{gt}(0)|V<v_{00}}^{-1}(q) = h_0\left(F_{U_{0}|G=g,V<v_{00}}^{-1}(q), t\right).$$
This implies that for all $y \in \Supp(Y_{g1}(0)|V<v_{00})$,
\begin{equation}
F_{Y_{g0}(0)|V<v_{00}}^{-1} \circ F_{Y_{g1}(0)|V<v_{00}}(y) = h_0(h_0^{-1}(y,1), 0).
    \label{eq:qq_1}
\end{equation}
By Assumption \ref{hyp:data1}, we have
\begin{eqnarray*}
    & & \Supp(Y_{010}) = \Supp(Y_{000}) \\
    & & \Rightarrow \Supp(Y_{10}(0)|V<v_{00}) = \Supp(Y_{00}(0)|V<v_{00}) \\
    & & \Rightarrow \Supp(h_0(U_0,0)|V<v_{00},G=1,T=0) = \Supp(h_0(U_0,0)|V<v_{00},G=0,T=0) \\
    & & \Rightarrow \Supp(U_0|V<v_{00},G=1) = \Supp(U_0|V<v_{00},G=0) \\
    & & \Rightarrow \Supp(h_0(U_0,1)|V<v_{00},G=1,T=1) = \Supp(h_0(U_0,1)|V<v_{00},G=0,T=1) \\
    & & \Rightarrow \Supp(Y_{11}(0)|V<v_{00}) = \Supp(Y_{01}(0)|V<v_{00}),
\end{eqnarray*}
where the third and fourth implications follow from Assumption \ref{hyp:monotonicity1}. Once combined with Equation \eqref{eq:qq_1}, the previous display implies that for all $y \in \Supp(Y_{11}(0)|V<v_{00})$,
$$F_{Y_{10}(0)|V<v_{00}}^{-1} \circ F_{Y_{11}(0)|V<v_{00}}(y) = F_{Y_{00}(0)|V<v_{00}}^{-1} \circ F_{Y_{01}(0)|V<v_{00}}(y).$$
This proves Equation (\ref{eq_2}), because $\{V<v_{00},G=g,T=0\}=\{D=0,G=g,T=0\}$.

\medskip
Finally, we show that
\begin{equation}
F_{Y_{01}(0)|V<v_{00}}=\lambda_{00} F_{001} + (1-\lambda_{00}) F_{Y_{01}(0)|S'}.
\label{eq_3}
\end{equation}
Suppose first that $\lambda_{00}\leq1$. Then, $v_{01}\leq v_{00}$ and $S'=\{V\in [v_{01},v_{00}),G=0\}$. Moreover, reasoning as for $P(S|G=1, V<v_{00})$, we get
\begin{eqnarray*}
\lambda_{00} & = & \frac{P(V<v_{01}|G=0)}{P(V<v_{00}|G=0)} = P(V<v_{01}|G=0, V< v_{00}), \\
F_{Y_{01}(0)|V<v_{00}}
& =&  \lambda_{00} F_{001} + (1-\lambda_{00})F_{Y_{01}(0)|S'}.
\end{eqnarray*}
If  $\lambda_{00}>1$, $v_{01}>v_{00}$ and $S'=\{V\in [v_{00},v_{01}),G=0\}$. We then have
\begin{eqnarray*}
1/\lambda_{00} & = & P(V<v_{00}|G=0, V< v_{01}),\\
F_{001} & =& 1/\lambda_{00} F_{Y_{01}(0)|V<v_{00}} + \left(1-1/\lambda_{00}\right)
F_{Y_{01}(0)|S'},
\end{eqnarray*}
so Equation (\ref{eq_3}) is also satisfied. \eqref{eq:F_CIC} follows by combining (\ref{eq_1}), (\ref{eq_2}) and (\ref{eq_3}).

\medskip
Now, under Assumption \ref{hyp:stable_control}, $\lambda_{00} = \lambda_{01}=1$. This and the
fact that $H_d \circ F_{d01} = F_{Q_d(Y_{d10})}$ shows that $F_{Y_{11}(d)|S}= F_{CIC,d}$.
This proves that $\tau_q=F^{-1}_{CIC,1}(q)-F^{-1}_{CIC,0}(q)$. Moreover,
\begin{eqnarray*}
W_{CIC} & =& \frac{p_{1|11} \int y dF_{Y_{111}}(y) - p_{1|10} \int y dF_{Q_1(Y_{101})}(y)}{p_{1|11}-p_{1|10}} \\
& -& \frac{p_{0|11} \int y dF_{Y_{011}}(y) - p_{0|10} \int y dF_{Q_0(Y_{001})}(y)}{p_{0|11}-p_{0|10}} \\
& =& \int y dF_{CIC,1}(y) - \int y dF_{CIC,0}(y) \\
& =&  E(Y_{11}(1) - Y_{11}(0)|S) \;_\Box
\end{eqnarray*}

\subsection*{Theorem \ref{thm:manygroups}}

We prove the first statement, the second and third following from similar arguments. Under the assumptions of the theorem, Assumptions \ref{hyp:timeinvariance2}-\ref{hyp:common_trends_within} and \ref{hyp:stable_control} are satisfied for the treatment and control groups $G^*=1$ and $G^*=0$.
Therefore, it follows from Theorem \ref{thm:IV-DID1} that
\begin{eqnarray}
&&W^*_{DID}(1,0)=E(Y(1)-Y(0)|S^*,G^*=1,T=1).\label{eq:multiple_groups7}
\end{eqnarray}
Similarly, one can show that
\begin{eqnarray}
&&W^*_{DID}(-1,0)=E(Y(1)-Y(0)|S^*, G^*=-1,T=1).\label{eq:multiple_groups8}
\end{eqnarray}
Moreover, by Assumption \ref{hyp:timeinvariance2} and $G\indep T$,
\begin{eqnarray*}
DID^*_D(1,0)P(G^*=1) &= &\left[E(D|G^*=1,T=1)-E(D|G^*=1,T=0)\right]P(G^*=1|T=1) \\
& = & P(S^*, G^*=1|T=1).
\end{eqnarray*}
Similarly, $DID^*_D(0,-1)P(G^*=-1)  = P(S^*,G^*=-1|T=1)$. Combining both equalities, we obtain
\begin{equation}\label{eq:multiple_groups9}
w_{10} = \frac{P(S^*, G^*=1|T=1)}{P(S^*, G^*=1|T=1)+P(S^*, G^*=-1|T=1)}=P(G^*=1|S^*,T=1).
\end{equation}
The result follows combining Equations \eqref{eq:multiple_groups7}-\eqref{eq:multiple_groups9} $_\Box$

\subsection*{Theorem \ref{thm:multivariate}}

We only prove the first statement, the second and third statements follow from similar arguments.

\medskip
$D_{01}\sim D_{00}$ and $D_{11}\gtrsim D_{10}$ combined with Assumption \ref{hyp:ordered}' imply that
\begin{eqnarray}
&& v^{d}_{01}= v^{d}_{00}, \text{ for every } d \in \{1,...,\overline{d}\}\label{eq:threshold2}\\
&& v^{d}_{11}\leq v^{d}_{10}, \text{ for every } d \in \{1,...,\overline{d}\}\label{eq:threshold3}.
\end{eqnarray}
Then, it follows from Assumption \ref{hyp:ordered}' and Equation \eqref{eq:threshold3}  that for every $d\in\{1,...,\overline{d}\}$,
\begin{eqnarray}\label{eq:change_treatment_rate_multi}
P(D_{11}\geq d)-P(D_{10}\geq d)&=&P(V\geq v^d_{11}|T=1,G=1)-P(V\geq v^d_{10}|T=0,G=1)\nonumber\\
&=&P(V\in [v^d_{11},v^d_{10})|G=1)\nonumber\\
&=&P(D(0) < d \leq D(1)|G=1).
\end{eqnarray}

\medskip
Then, for every $g \in \{0,1\}$,
\begin{eqnarray}\label{eq:proof_multi_WDID1}
&&E(Y_{g1})-E(Y_{g0})\nonumber\\
&=&\sum_{d=0}^{\overline{d}}E(Y_{g1}(d)|G=g,V\in [v^d_{g1},v^{d+1}_{g1}))P(V\in [v^d_{g1},v^{d+1}_{g1})|G=g)\nonumber\\
&-&\sum_{d=0}^{\overline{d}}E(Y_{g0}(d)|G=g,V\in [v^{d}_{g0},v^{d+1}_{g0}))P(V\in [v^{d}_{g0},v^{d+1}_{g0})|G=g)\nonumber\\
&=&\sum_{d=1}^{\overline{d}}E(Y_{g1}(d)-Y_{g1}(d-1)|G=g,V\in [v^d_{g1},v^{d}_{g0}))P(V\in [v^d_{g1},v^{d}_{g0})|G=g)\nonumber\\
&+&\sum_{d=0}^{\overline{d}}E(Y_{g1}(d)-Y_{g0}(d)|G=g,V\in [v^{d}_{g0},v^{d+1}_{g0}))P(V\in [v^{d}_{g0},v^{d+1}_{g0})|G=g)\nonumber\\
&=&\sum_{d=1}^{\overline{d}}E(Y_{g1}(d)-Y_{g1}(d-1)|D(0) < d \leq D(1))P(D(0) < d \leq D(1)|G=g)\nonumber\\
&+&E(Y_{g1}(0))-E(Y_{g0}(0)).
\end{eqnarray}
The first, second and third equalities respectively follow from Assumption \ref{hyp:ordered}', Equations \eqref{eq:threshold2} and \eqref{eq:threshold3} combined
with Assumption \ref{hyp:ordered}', and Assumptions \ref{hyp:ordered}' and \ref{hyp:common_trends_within}.

\medskip
Combining Equation \eqref{eq:proof_multi_WDID1} with Equation \eqref{eq:threshold2} and Assumption \ref{hyp:common_trends_between} imply that
\begin{align*}
DID_Y =& \sum_{d=1}^{\overline{d}}E(Y_{11}(d)-Y_{11}(d-1)|D(0) < d \leq D(1))P(D(0) < d \leq D(1)|G=1).
\end{align*}
The result follows from Equation \eqref{eq:change_treatment_rate_multi}, after dividing each side of the previous display by $DID_D$ $_\Box$

\subsection*{Proof of Theorem \ref{thm:partial_ident}}

\textbf{Proof of 1}

\medskip
We only prove that $\underline{W}_{TC}$ is a lower bound when $\lambda_{00}>1$. The proofs for the upper bound and when $\lambda_{00}<1$ are symmetric.

\medskip
We have
\begin{eqnarray*}
&&E(Y_{11}(1)-Y_{11}(0)|S)P(S|G=1)\nonumber\\
&=&E(Y_{11})-E(Y_{10})-E(Y_{11}(1)-Y_{10}(1)|V\geq v_{00})P(V\geq v_{00}|G=1)\nonumber\\
&-&E(Y_{11}(0)-Y_{10}(0)|V< v_{00})P(V< v_{00}|G=1)\\
&=&E(Y_{11})-E(Y_{10}) - E(Y_{01}(1)-Y_{00}(1)|V\geq v_{00})p_{1|10}\nonumber\\
&-&E(Y_{01}(0)-Y_{00}(0)|V< v_{00})p_{0|10}\\
&=&E(Y_{11})-E(Y_{10}) - \left(E(Y_{01}(1)|V\geq v_{00})-E(Y_{100})\right)p_{1|10}\nonumber\\
&-&\left(E(Y_{01}(0)|V< v_{00})-E(Y_{000})\right)p_{0|10}.
\end{eqnarray*}
The first equality follows from Equation \eqref{eq:proof_WTC1}, the second from Assumptions \ref{hyp:timeinvariance2} and \ref{hyp:common_trends_conditional}'.
Thus, the proof will be complete if we can show that $\overline{\delta}_1$
and $\overline{\delta}_0$ are respectively upper bounds for $E(Y_{01}(1)|V\geq v_{00})-E(Y_{100})$ and $E(Y_{01}(0)|V< v_{00})-E(Y_{000})$.

\medskip
When $\lambda_{00}>1$, Assumption \ref{hyp:timeinvariance2} implies that $v_{00}<v_{01}$. Then,
reasoning as in the proof of Theorem \ref{thm:ident_CIC}, we obtain
$P(V\geq v_{01}|G=0, T=1, V\geq v_{00}) = \lambda_{01}$ and	
\begin{eqnarray*}
E(Y_{01}(1)|V\geq v_{00})&=&\lambda_{01}E(Y_{01}(1)|V\geq v_{01})+(1-\lambda_{01})E(Y_{01}(1)|S')\nonumber\\
&\leq &\lambda_{01}E(Y_{101})+(1-\lambda_{01})\overline{y} = \int y d\underline{F}_{101}(y).
\end{eqnarray*}
This proves that $\overline{\delta}_1$ is an upper bound for $E(Y_{01}(1)|V\geq v_{00})-E(Y_{100})$.

\medskip
Similarly, $P(V< v_{00}|G=0, T=1, V< v_{01}) = 1/\lambda_{00}$ and
\begin{eqnarray*}
F_{Y_{001}}&=&1/\lambda_{00}F_{Y_{01}(0)|V< v_{00}}+\left(1-1/\lambda_{00}\right)F_{Y_{01}(0)|S'}.
\end{eqnarray*}
Following \cite{Horowitz95}, the last display implies that
\begin{eqnarray*}
E(Y_{01}(0)|V< v_{00})&\leq& \int y d\underline{F}_{001}(y).
\end{eqnarray*}
This proves that $\overline{\delta}_0$ is an upper bound for $E(Y_{01}(0)|V<v_{00})-E(Y_{000})$.

\medskip
\textbf{Proof of 2}

\medskip
\textbf{Construction of the bounds.}

\medskip
We only establish the validity of the bounds for $F_{Y_{11}(0)|S}(y)$. The reasoning is similar for $F_{Y_{11}(1)|S}(y)$. Bounds for $\Delta$ and $\tau_q$ directly follow from those for the cdfs. Hereafter, to simplify the notation, we let $T_0=F_{Y_{01}(0)|S'}$. Following the notation introduced in Subsection \ref{sec:partial} combined with Equations \eqref{eq:F_CIC} and \eqref{eq_3} , we then have  $G_0(T_0)=F_{Y_{01}(0)|V<v_{00}}$ and $C_0(T_0)=F_{Y_{11}(0)|S}$.

\medskip
We start considering the case where $\lambda_{00}<1$. We first show that in such instances, $0\leq T_0, G_0(T_0), C_0(T_0) \leq 1$ if and only if
\begin{equation}
\underline{T_0}\leq T_0 \leq \overline{T_0}.
\label{eq_7}
\end{equation}
$G_0(T_0)$ is included between 0 and 1 if and only if
$$\frac{-\lambda_{00} F_{001}}{1-\lambda_{00}} \leq T_0 \leq \frac{1-\lambda_{00} F_{001}}{1-\lambda_{00}},$$
while $C_0(T_0)$ is included between 0 and 1 if and only if
$$\frac{H_0^{-1}(\lambda_{10}F_{011})-\lambda_{00} F_{001}}{1-\lambda_{00}} \leq T_0 \leq \frac{H_0^{-1}(\lambda_{10}F_{011}+(1-\lambda_{10}))-\lambda_{00} F_{001}}{1-\lambda_{00}}.$$
Since $-\lambda_{00} F_{001}/(1-\lambda_{00})\leq 0$ and $(1-\lambda_{00}
F_{001})/(1-\lambda_{00})\geq 1$, $T_0$, $G_0(T_0)$ and $C_0(T_0)$ are all included between 0 and 1 if and only if
\begin{equation}
M_0\left(\frac{H_0^{-1}(\lambda_{10}F_{011})-\lambda_{00} F_{001}}{1-\lambda_{00}}\right)\leq T_0 \leq m_1\left(\frac{H_0^{-1}(\lambda_{10}F_{011}+(1-\lambda_{10}))-\lambda_{00} F_{001}}{1-\lambda_{00}}\right),
\label{eq_7bis}
\end{equation}
where $M_0(x)=\max(0,x)$ and $m_1(x)=\min(1,x)$. Composing each term of these inequalities by $M_0(.)$ and then by $m_1(.)$ yields Equation \eqref{eq_7},
since $M_0(T_0)=m_1(T_0)=T_0$ and $M_0 \circ m_1=m_1 \circ M_0$.

\medskip
Now, when $\lambda_{00}<1$, $G_0(T_0)$ is increasing in $T_0$, so $C_0(T_0)$ as well is increasing in
$T_0$. Combining this with \eqref{eq_7} implies that for every $y'$,
\begin{equation}
C_0(\underline{T_0})(y')\leq C_0(T_0)(y') \leq C_0(\overline{T_0})(y').
\label{eq_8}
\end{equation}
Because $C_0(T_0)(y)$ is a cdf,
\begin{equation*}
C_0(T_0)(y) = \inf_{y'\geq y} C_0(T_0)(y') \leq \inf_{y'\geq y} C_0(\overline{T_0})(y')=\overline{F}_{CIC,0}(y).
\end{equation*}
This proves the result for the upper bound. The result for the lower bound follows similarly.

\medskip
Let us now turn to the case where $\lambda_{00}>1$. Using the same reasoning as above, we get that $G_0(T_0)$ and $C_0(T_0)$ are included between $0$ and $1$ if and only if
\begin{eqnarray*}
  \frac{\lambda_{00} F_{001}-1}{\lambda_{00}-1} & \leq T_0 \leq & \frac{\lambda_{00} F_{001}}{\lambda_{00}-1}, \\
  \frac{\lambda_{00} F_{001}-H_0^{-1}(\lambda_{10}F_{011}+(1-\lambda_{10}))}{\lambda_{00}-1}& \leq T_0 \leq &
  \frac{\lambda_{00} F_{001}-H_0^{-1}(\lambda_{10}F_{011})}{\lambda_{00}-1}.
\end{eqnarray*}
The inequalities in the first line are not binding since they are implied by those on the second
line. Thus, we also get \eqref{eq_7bis}. Hence, $0\leq T_0, G_0(T_0),
C_0(T_0) \leq 1$ if and only if
\begin{equation}
\overline{T_0}\leq T_0 \leq \underline{T_0}.
\label{eq_9}
\end{equation}
Besides, when $\lambda_{00}>1$, $G_0(T_0)$ is decreasing in $T_0$, so $C_0(T_0)$ is also
decreasing in $T_0$. Combining this with Equation \eqref{eq_9} implies that for every $y$, Equation \eqref{eq_8}
holds as well. This proves the result.

\medskip
\textbf{Sketch of the proof of sharpness.}

\medskip
The full proof is in the supplementary material (see \citep{deChaisemartin15}). We only consider the sharpness of $\underline{F}_{CIC,0}$, the reasoning being similar for the upper bound. The proof is also similar and actually simpler for $d=1$. The corresponding bounds are proper cdfs, so we do not have to consider converging sequences of cdfs as we do in case b) below.

\medskip
\textbf{a. $\lambda_{00}>1$.} We show that if Assumptions \ref{hyp:data1}-\ref{hyp:increasingbounds} hold,  then $\underline{F}_{CIC,0}$ is sharp.
For that purpose, we construct $\widetilde{h}_0, \widetilde{U}_0, \widetilde{V}$ such that:
\begin{enumerate}
    \item[(i)] $Y=\widetilde{h}_0(\widetilde{U}_0,T)$ when $D=0$ and $D=1\{\widetilde{V}\geq v_{GT}\}$;
    \item[(ii)] $(\widetilde{U}_0, \widetilde{V}) \indep T |G$;
    \item[(iii)] $\widetilde{h}_0(.,t)$ is strictly increasing for $t\in \{0,1\}$;
    \item[(iv)] $F_{\widetilde{h}_0(\widetilde{U}_0, 1)|G=0,T=1,\widetilde{V}\in [v_{00},v_{01})}=\underline{T}_0$.
\end{enumerate}
Because we can always define $\widetilde{Y}(0)$ as $\widetilde{h}_0(\widetilde{U}_0,T)$ when $D=1$ without contradicting the data,
(i)-(iii) ensures that Assumptions \ref{hyp:timeinvariance2}  and \ref{hyp:monotonicity1} (for $d=0$) are satisfied with $\widetilde{h}_0, \widetilde{U}_0$ and $\widetilde{V}$. (iv) ensures that the DGP corresponding to $(\widetilde{h}_0, \widetilde{U}_0, \widetilde{V})$ rationalizes the bound.

\medskip
The construction of $\widetilde{h}_0$, $\widetilde{U}_0$, and $\widetilde{V}$ is long, so its presentation is deferred to the supplementary material.

\medskip
\textbf{b. $\lambda_{00}<1$.} The idea is similar as in the previous case. A difference, however, is that when
$\lambda_{00}<1$, $\underline{T}_0$ is not a proper cdf, but a defective one, since $\lim_{y\rightarrow
\overline{y}} \underline{T}_0(y)<1$. As a result, we cannot define a DGP such that
$\widetilde{T}_0=\underline{T}_0$, However, by Lemma S\ref{lem:suite_Tn}, there exists a sequence
$(\underline{T}^k_0)_{k\in \mathbb{N}}$ of cdfs such that $\underline{T}_0^k \rightarrow \underline{T}_0$,
$G_0(\underline{T}^{k}_0)$ is an increasing bijection from $\Supp(Y)$ to $(0,1)$ and
$C_0(\underline{T}^{k}_0)$ is increasing and onto $(0,1)$. We can then construct a sequence of DGP
$(\widetilde{h}^k_0(.,0),\widetilde{h}^k_0(.,1),\widetilde{U}^k_0,\widetilde{V}^k)$ such that Points (i)
to (iii) listed above hold for every $k$, and such that $\widetilde{T}_0^k=\underline{T}^k_0$. Since
$\underline{T}^{k}_0(y)$ converges to $\underline{T}_0(y)$ for every $y$ in $\interior{\Supp(Y)}$, we
thus define a sequence of DGP such that $\widetilde{T}_0^k$ can be arbitrarily close to
$\underline{T}_0$ on $\interior{\Supp(Y)}$ for sufficiently large $k$. Since $C_0(.)$ is continuous, this
proves that $\underline{F}_{CIC,0}$ is sharp on $\interior{\Supp(Y)}$. Again, this construction is long and its exposition is deferred to the supplementary material $_\Box$

\subsection*{Theorem \ref{thm:as_nor_point}}

\textbf{Proof of 1 and 2}

\medskip
Asymptotic normality is obvious by the central limit theorem and the delta method. Consistency of the bootstrap follows by consistency of the bootstrap for sample means (see, e.g., \citep{vanderVaart00}, Theorem 23.4) and the delta method for bootstrap (\citep{vanderVaart00}, Theorem 23.5). A convenient way to obtain the asymptotic variance is to use repeatedly the following argument. If
$$\sqrt{n}\left(\widehat{A}-A\right)=\frac{1}{\sqrt{n}}\Sig a_i + o_P(1) \text{ and } \; \sqrt{n}\left(\widehat{B}-B\right)=\frac{1}{\sqrt{n}}\Sig b_i + o_P(1),$$
then Lemma S\ref{lem:ratio} ensures that
\begin{equation}
\label{eq:lin_ratio}
\sqrt{n}\left(\frac{\widehat{A}}{\widehat{B}}-\frac{A}{B}\right) = \frac{1}{\sqrt{n}}\Sig
\frac{a_i - (A/B) b_i}{B} + o_P(1).
\end{equation}
This implies for instance that
$$\sqrt{n}\left(\widehat{E}(Y_{11}) - E(Y_{11})\right)=  \frac{1}{\sqrt{n}}\Sig \frac{G_i T_i (Y_i - E(Y_{11}))}{p_{11}}+ o_P(1),$$
and similarly for $\widehat{E}(D_{11})$. Applying repeatedly this argument, we obtain, after some algebra,
$$\sqrt{n}\left(\widehat{W}_{DID} - \Delta\right)  = \frac{1}{\sqrt{n}}\Sig  \psi_{DID,i} + o_P(1),$$
where, omitting the index $i$, $\psi_{DID}$ is defined by
\begin{align}
\psi_{DID}= \frac{1}{DID_D} & \left[\frac{GT(\eps-E(\eps_{11}))}{p_{11}}-
\frac{G(1-T)(\eps-E(\eps_{10}))}{p_{10}} - \frac{(1-G)T(\eps-E(\eps_{01}))}{p_{01}} \right. \nonumber\\
& \left. \; + \frac{(1-G)(1-T)(\eps-E(\eps_{00}))}{p_{00}}\right]
\label{eq:psi_DID}
\end{align}
and $\eps = Y - \Delta D$. Similarly,
$$\sqrt{n}\left(\widehat{W}_{TC} - \Delta\right) = \frac{1}{\sqrt{n}}\Sig  \psi_{TC,i} + o_P(1),$$
where $\psi_{TC}$ is defined by
\begin{align}
\psi_{TC}  = & \frac{1}{E(D_{11})-E(D_{10})}  \bigg\{ \frac{GT(\eps-E(\eps_{11}))}{p_{11}}-
\frac{G(1-T)(\eps + (\delta_1-\delta_0) D - E(\eps_{10} + (\delta_1-\delta_0) D_{10}))}{p_{10}} \nonumber \\
 &   - E(D_{10}) D (1-G)\left[\frac{T(Y-E(Y_{101}))}{p_{101}} - \frac{(1-T)(Y-E(Y_{100}))}{p_{100}}\right]  \nonumber \\
 & - (1-E(D_{10}))(1-D)(1-G) \left[\frac{T(Y-E(Y_{001}))}{p_{001}} - \frac{(1-T)(Y-E(Y_{000}))}{p_{000}}\right]  \bigg\}.
\label{eq:psi_TC}
\end{align}

\medskip
\textbf{Proof of 3}

\medskip
We first show that $(\widehat{F}_{Y_{11}(0)|S}, \widehat{F}_{Y_{11}(1)|S})$ tends to a continuous
gaussian process. Let $\widetilde{\theta} = (F_{000},
F_{001},...,F_{111},\lambda_{10}, \lambda_{11})$. By Lemma S\ref{lem:conv_distr_base}, $\widehat{\widetilde{\theta}}= (\widehat{F}_{000},
\widehat{F}_{001},...,\widehat{F}_{111},\widehat{\lambda}_{10}, \widehat{\lambda}_{11})$ converges to a
continuous gaussian process. Let
$$\pi_d: (F_{000}, F_{001},...,F_{111},\lambda_{10}, \lambda_{11}) \mapsto
\left(F_{d10},F_{d00},F_{d01},F_{d11},1,\lambda_{1d}\right), \quad d \in \{0,1\},$$ so that
$(\widehat{F}_{Y_{11}(0)|S}, \widehat{F}_{Y_{11}(1)|S}) = \left(R_1 \circ
\pi_0(\widetilde{\theta}), R_1\circ \pi_1(\widetilde{\theta})\right)$, where $R_1$ is defined as in
Lemma S\ref{lem:had_diff}. $\pi_d$ is Hadamard differentiable as a linear continuous map. Because
$F_{d10},F_{d00},F_{d01},F_{d11}$ are continuously differentiable with strictly positive derivative
by Assumption \ref{hyp:techconditioninference1}, $\lambda_{1d} >0$, and $\lambda_{1d} \ne 1$ under Assumption \ref{hyp:data1}, $R_1$ is also Hadamard differentiable at
$(F_{d10},F_{d00},F_{d01},F_{d11},1,\lambda_{1d})$ tangentially to $(\mathcal{C}^0)^4\times \R^2$. By the functional delta
method (see, e.g., \citep{vanderVaart96}, Lemma 3.9.4), $(\widehat{F}_{Y_{11}(0)|S},
\widehat{F}_{Y_{11}(1)|S})$ tends to a continuous gaussian process.

\medskip
Now, by integration by parts for Lebesgue-Stieljes integrals,
$$\Delta = \int_{\underline{y}}^{\overline{y}} F_{Y_{11}(0)|S}(y)-F_{Y_{11}(1)|S}(y) dy.$$
Moreover, the map $\varphi_1:(F_1,F_2) \mapsto \int_{\Supp(Y)} (F_2(y)-F_1(y)) dy $, defined on the
domain of bounded c\`adl\`ag functions, is linear. Because $\Supp(Y)$ is bounded by Assumption \ref{hyp:techconditioninference1},
$\varphi_1$ is also continuous with respect to the supremum norm. It is thus Hadamard
differentiable. Because $\widehat{\Delta} =
\varphi_1\left(\widehat{F}_{Y_{11}(1)|S},\widehat{F}_{Y_{11}(0)|S}\right)$, $\widehat{\Delta} $ is
asymptotically normal by the functional delta method. The asymptotic normality of
$\widehat{\tau}_q$ follows along similar lines. By Assumption \ref{hyp:techconditioninference1}, $F_{Y_{11}(d)|S}$ is differentiable with
strictly positive derivative on its support. Thus, the map $(F_1,F_2) \mapsto F_2^{-1}(q)-
F_1^{-1}(q)$ is Hadamard differentiable at $(F_{Y_{11}(0)|S},F_{Y_{11}(1)|S})$ tangentially to the
set of functions that are continuous at $(F_{Y_{11}(0)|S}^{-1}(q),F_{Y_{11}(1)|S}^{-1}(q))$ (see
Lemma 21.3 in \citep{vanderVaart00}). By the functional delta method, $\widehat{\tau}_q$ is
asymptotically normal.

\medskip
The validity of the bootstrap follows along the same lines. By Lemma S\ref{lem:conv_distr_base}, the
bootstrap is consistent for $\widehat{\widetilde{\theta}}$. Because both the LATE and LQTE are Hadamard
differentiable functions of $\widehat{\widetilde{\theta}}$, as shown above, the result simply follows by the
functional delta method for the bootstrap (see, e.g., \citep{vanderVaart00}, Theorem 23.9).

\medskip
Finally, we compute the asymptotic variance of both estimators. The functional delta method also implies that both estimators are asymptotically linear. To compute their asymptotic variance, it suffices to provide their asymptotic linear approximation. For that purpose, let us first linearize $F_{Y_{11}(d)|S}(y)$, for all $y$. It follows from the proof of the first point of Lemma S\ref{lem:had_diff} that the mapping $\phi_1:(F_1,F_2,F_3) \mapsto F_1 \circ F_2^{-1} \circ F_3$ is Hadamard differentiable at $(F_{d10}, F_{d00}, F_{d01})$, tangentially to $(\mathcal{C}^0)^3$. Moreover applying the chain rule, we obtain
$$d\phi_1(h_1,h_2,h_3) = h_1\circ  Q_d^{-1} + H_d' \circ F_{d01} \times \left[-h_2 \circ Q_d^{-1} + h_3\right].$$
Applied to $(F_1,F_2,F_3) = (F_{d10}, F_{d00}, F_{d01})$, this and  the functional delta method once more imply that
$$\sqrt{n}\left(\widehat{H}_d \circ \widehat{F}_{d01} - H_d\circ F_{d01}\right) = d\phi_1(h_{1n},h_{2n},h_{3n}) + o_P(1),$$
where the $o_P(1)$ is uniform over $y$ and $h_{1n}=\sqrt{n}(\widehat{F}_{d10}-F_{d10})$. $h_{2n}$ and $h_{3n}$ are defined similarly. Furthermore, applying Lemma S\ref{lem:ratio} yields, uniformly over $y$,
$$h_{1n}(y) = \frac{1}{\sqrt{n}}\Sig \frac{\mathds{1}\{D_i=d\}G_i(1-T_i)(\mathds{1}\{Y_i \leq y\}-F_{d10}(y))}{p_{d10}} + o_P(1).$$
A similar expression holds for $h_{2n}$ and $h_{3n}$. Hence, by continuity of $d\phi_1$, we obtain, after some algebra,
\begin{align*}
& \sqrt{n}\left(\widehat{H}_d \circ \widehat{F}_{d01}(y) - H_d\circ F_{d01}(y)\right) \\
= & \frac{1}{\sqrt{n}}\Sig \mathds{1}\{D_i=d\}\left\{\frac{G_i(1-T_i)(\mathds{1}\{Q_d(Y_i) \leq y\}-H_d\circ F_{d01}(y))}{p_{d10}} +  (1-G_i) H'_d\circ F_{d01}(y) \right. \\
& \qquad \left. \times \left[ - \frac{(1-T_i)(\mathds{1}\{Q_d(Y_i) \leq y\}-F_{d01}(y))}{p_{d00}} + \frac{T_i(\mathds{1}\{Y_i \leq y\}-F_{d01}(y))}{p_{d01}} \right]\right\} +  o_P(1),
\end{align*}
which holds uniformly over $y$. Applying repeatedly Lemma S\ref{lem:ratio}, we then obtain, after some algebra,
$$\sqrt{n}\left(\widehat{F}_{Y_{11}(d)|S}(y) - F_{Y_{11}(d)|S}(y)\right) = \frac{1}{\sqrt{n}}\Sig \Psi_{di}(y) + o_P(1),$$
where, omitting the index $i$,

{\small \begin{align*}
\Psi_{d}(y)  & = \frac{1}{p_{d|11}-p_{d|10}} \left\{\frac{G T}{p_{11}} \left[\mathds{1}\{D=d\}\mathds{1}\{Y \leq y\} - p_{d|11}F_{d11}(y) -
F_{Y_{11}(d)|S}(y)\left(\mathds{1}\{D=d\} - p_{d|11}\right)\right] \right. \\
& + \frac{G (1-T)}{p_{10}} \left[-\mathds{1}\{D=d\}\left(\mathds{1}\{Q_d(Y) \leq y\}-H_d\circ F_{d01}(y)\right) +
\left(\mathds{1}\{D=d\} - p_{d|10}\right) \left(F_{Y_{11}(d)|S}(y) - H_d \circ F_{d01}(y)\right)\right] \\
& \left. + p_{d|10} (1-G)\mathds{1}\{D=d\}H'_d\circ F_{d01}(y) \left[\frac{(1-T)(\mathds{1}\{Q_d(Y) \leq y\}-F_{d01}(y))}{p_{d00}} -
\frac{T(\mathds{1}\{Y \leq y\}-F_{d01}(y))}{p_{d01}} \right]\right\}.
\end{align*}}

By the functional delta method, this implies that we can also linearize $\widehat{W}_{CIC}$ and $\widehat{\tau}_q$. Moreover, we obtain by the chain rule the following influence functions:
\begin{align}
\psi_{CIC}  = & \int \Psi_0(y) -\Psi_1(y)dy,    \label{eq:psi_W_CIC} \\
\psi_{q,CIC}  = & \left[\frac{\Psi_1}{f_{Y_{11}(1)|S}}\right]\circ F^{-1}_{Y_{11}(1)|S}(q) -
\left[\frac{\Psi_0}{f_{Y_{11}(0)|S}}\right]\circ F^{-1}_{Y_{11}(0)|S}(q). \label{eq:psi_tau_CIC}
\end{align}

\end{document}